\pdfoutput=1

\documentclass[11pt]{article}

\usepackage{acl}
\usepackage{times}
\usepackage{latexsym}

\usepackage[T1]{fontenc}

\usepackage[utf8]{inputenc}

\usepackage{microtype}

\usepackage{inconsolata}

\usepackage{graphicx}

\usepackage{multirow}
\usepackage{bbding}
\usepackage{algorithm}
\usepackage{algorithmic}

\title{ODTQA-FoRe: An Open-Domain Tabular Question Answering Dataset for Future Data Forecasting and Reasoning}

\author{
  \textbf{Zhensheng Wang\textsuperscript{1,2}},
  \textbf{Xiaole Liu\textsuperscript{3}},
  \textbf{Wenmian Yang\textsuperscript{2,\textdagger}},
  \textbf{Kun Zhou\textsuperscript{1,2}},
  \textbf{Yiquan Zhang\textsuperscript{2}},
  \textbf{Weijia Jia\textsuperscript{2,4,\textdagger}}
  \\
  \textsuperscript{1}School of Artificial Intelligence, Beijing Normal University, Beijing, PR China\\
  \textsuperscript{2}Institute of Artificial Intelligence and Future Networks, Beijing Normal University, Zhuhai, PR China\\
  \textsuperscript{3}Faculty of Arts and Sciences, Beijing Normal University, Zhuhai, PR China\\
  \textsuperscript{4}Beijing Normal-Hong Kong Baptist University, Zhuhai, PR China
  \\
  \small{
    \{jensenwang, xiaoleliu, zhoukun\}@mail.bnu.edu.cn, 
    \{wenmianyang, jiawj\}@bnu.edu.cn, 
    zhangyq987@hotmail.com
  }
}

\begin{document}
\maketitle

\begingroup
\renewcommand\thefootnote{\textdagger}
\footnotetext{Corresponding authors.}
\endgroup

\begin{abstract}
The rapid development of LLMs has significantly advanced tabular question answering, but most systems cannot perform future-oriented numerical prediction. To address this gap, we introduce a novel task, Open-Domain Tabular Question Answering for Future Data Forecasting and Reasoning, and propose the first dataset to cover time-series forecasting and forecast-based reasoning scenarios using real estate data. This task poses challenges in retrieving precise historical data, overcoming the forecasting limitations of LLMs, and standardizing responses for diverse queries. To solve the above challenges, we propose TimeFore, an LLM agent-based framework that decomposes the problem into three collaborative roles: a Retriever autonomously generates SQL to fetch data, a Forecaster invokes external time-series models for higher accuracy, and an Analyzer synthesizes the results to construct a precise and consistent final answer. Extensive experiments demonstrate the effectiveness of our TimeFore. The dataset and code are available at \url{https://github.com/jensenw1/ODTQA-FoRe}.

\end{abstract}

\section{Introduction}

With the rapid development of large language models (LLMs), question-answering systems built upon these models have achieved remarkable progress across a wide range of QA tasks \cite{openQA2023nips}. 
In particular, the integration of Retrieval-Augmented Generation (RAG) \cite{RAG2020nips} techniques has demonstrated powerful cross-task generalization capabilities, creating new opportunities and breakthroughs for open-domain QA systems \cite{ChenFWB2017ACL,OpenTab@Kong0SSLFRK24,chen-etal-2025-llm-based}. As user demands evolve, there is growing expectation for LLMs not only to accurately answer traditional knowledge-based questions, but also to exhibit enhanced numerical reasoning abilities \cite{TabLaP@aaai2025Wang,DBLP:journals/fcsc/ZhangWDZC25}. As a result, considerable research attention has shifted toward open-domain numerical reasoning over tables, a direction commonly known as open-domain tabular question answering (ODTQA) \cite{herzig2021open,openwikitable@acl/KweonKCJC23}. ODTQA methods typically leverage RAG techniques, as well as the models’ abilities in program synthesis and database query generation. These methods have proven particularly effective for numerical reasoning in vertical domains including real estate, healthcare, and finance.

\begin{figure}[t]
 \centering
 \includegraphics[width=\columnwidth]{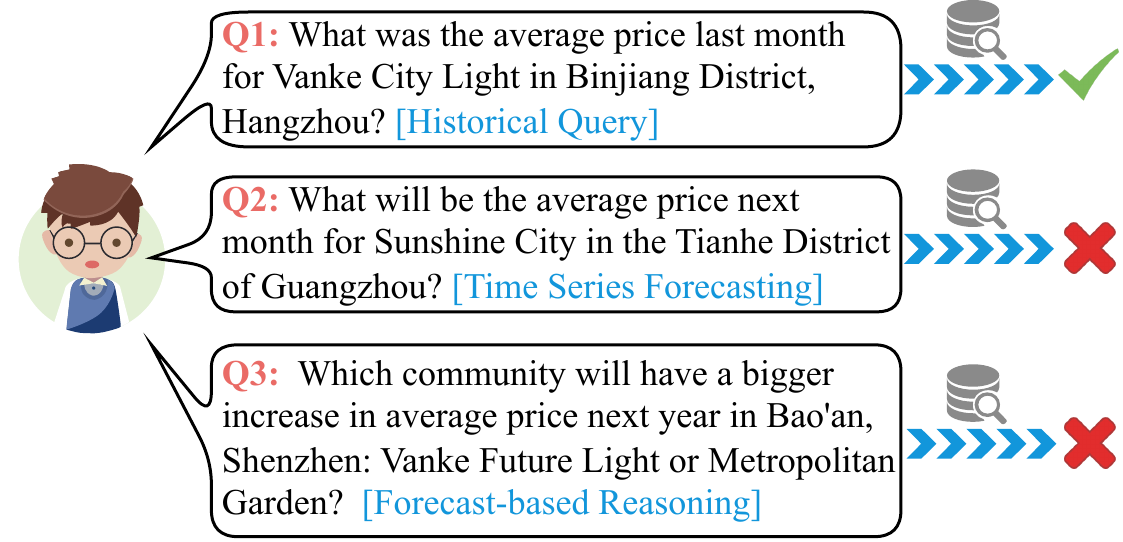}
 \centering
 \caption{Example questions of future data forecasting and reasoning.}
 \label{fig:Query_type}
\end{figure}

However, current open-domain tabular QA studies predominantly focus on retrieval and reasoning based on data from static databases, with limited attention to questions requiring future data predictions, as shown in Figure \ref{fig:Query_type}. In practice, users frequently ask questions related to the forecasting and reasoning of future data, such as forecasting housing prices in the coming months (Figure \ref{fig:Query_type}, Q2) or comparing price trends across different communities (Figure \ref{fig:Query_type}, Q3), to inform investment and financial decision-making. While existing methods face significant challenges in addressing these queries for two main reasons. First, LLMs are primarily trained on static textual data and lack dedicated mechanisms for modeling dynamic time-series data, which limits their ability to make effective future predictions \cite{mingtian@nips/TanMGAH24,ye-etal-2025-optimizing}. Second, in open-domain scenarios, users typically cannot provide continuous and accurate historical data to support forecasting. Consequently, systems must autonomously generate query statements (e.g., SQL) to retrieve relevant historical data from large databases for analysis and prediction.

To address these issues, we propose a novel research task, namely Open-Domain Tabular Question Answering for Future Data Forecasting and Reasoning (ODTQA-FoRe), aiming to expand the scope of LLM-based QA systems into future data prediction. This task requires the QA system to possess cross-domain data retrieval capabilities, future-oriented time series forecasting skills, and comprehensive table data understanding and reasoning abilities to meet deeper user needs. However, this research direction is still in its infancy, lacking relevant datasets and benchmark frameworks, which restricts effective algorithm development and evaluation. To bridge this gap, we introduce the first Open-Domain Tabular Question Answering Dataset, ODTQA-FoRe, constructed using real estate data and incorporating both time series forecasting and forecast-based reasoning questions, thus addressing the absence of relevant research resources in this area.

Specifically, ODTQA-FoRe comprises 28,507 QA pairs constructed from real estate transaction records spanning 2022–2024 across 10 Chinese cities. Following the open-domain paradigm, the system must autonomously identify and retrieve relevant tables from a candidate pool of 288 tables (averaging 845 rows each), rather than relying on a pre-specified target table. We restrict the dataset to the real estate vertical to ensure dense and continuous temporal data for reliable forecasting evaluation. Despite this vertical focus, the task formulation remains generic, rendering the proposed framework readily extensible to other domains. The dataset contains 8,042 time-series forecasting questions and 20,465 forecast-based reasoning questions, posing substantial challenges for QA systems.


This new task presents three distinct challenges: First, the challenge of historical data retrieval, where systems must accurately identify relevant tables from massive databases and retrieve pertinent historical data sequences based on user queries, demanding significantly higher retrieval precision compared to traditional tabular QA tasks. Second, the challenge of forecasting future data from historical sequences, as LLMs inherently exhibit limited predictive abilities, necessitating the development of methods to overcome these prediction limitations. Third, the challenge of problem type identification and standardized response generation, as users pose diverse question types (e.g., time series forecasting and forecast-based reasoning), each requiring different response formats and prompting methods. Thus, it is crucial for the system to accurately distinguish between question types and generate standardized responses.

To address the challenges of ODTQA-FoRe, we present TimeFore, an LLM agent-based framework that aims to establish a foundational baseline in this unique area. The framework decomposes the problem into three collaborative roles: a Retriever that tackles the data acquisition challenge by identifying relevant tables and autonomously generating SQL queries to fetch historical data; a Forecaster that addresses the LLM's inherent predictive weaknesses by invoking external time-series models to enhance forecasting accuracy; and an Analyzer that synthesizes the retrieved data and forecasts into a standardized response format. As ODTQA-FoRe is our newly proposed benchmark without directly comparable baselines, TimeFore aims to set a performance standard for ODTQA-FoRe, further underscoring its importance in advancing this field.

\begin{table*}[htbp]\centering\small
\begin{tabular}{ccccccc}
\hline
Dataset           & Open Domain  & \# of QA pairs & \# Tables & Answer format & Multi-table     & TS forecasting \\ \hline
WikiTableQuestion & \XSolidBrush & 22033         & 2108     & Text          & \XSolidBrush  & \XSolidBrush  \\
Spider            & \XSolidBrush & 10181         & 1020     & SQL           & \Checkmark      & \XSolidBrush  \\
Open-WikiTable    & \Checkmark   & 67023         & 24680    & Text,SQL      & \XSolidBrush  & \XSolidBrush  \\
NQ-TABLES         & \Checkmark   & 11628         & 169898   & Text          & \XSolidBrush   & \XSolidBrush  \\
RETQA             & \Checkmark   & 20762         & 4932     & Text,SQL      & \Checkmark     & \XSolidBrush  \\ \hline
ODTQA-FoRe       & \Checkmark   & 28507         & 288      & Text,SQL      & \Checkmark     & \Checkmark    \\ \hline
\end{tabular}
\caption{Comparison with existing datasets. ``Open Domain'' indicates table retrieval from a large corpus, whereas ``Closed Domain'' (marked with `\XSolidBrush') specifies QA on given tables. “TS forecasting” refers to time-series forecasting.}
\label{Dataset_Comparison}
\end{table*}

The main contributions of this paper are as follows:

\begin{itemize}
 \item We define the novel task of Open-Domain Tabular Question Answering for Future Data Forecasting and Reasoning. To support this, we construct a new benchmark dataset using real-world real estate data, incorporating diverse question types such as forecasting and forecast-based reasoning, filling a critical gap in the field.

 \item We propose TimeFore as the benchmark framework, effectively addressing the key challenges of historical data retrieval, future data prediction, and question type distinction and standardized response generation through its three dedicated agents.


 \item We conduct comprehensive experiments to validate TimeFore's effectiveness and establish strong performance benchmarks across five mainstream LLMs. Furthermore, our detailed ablation studies dissect the framework's components, identify key performance bottlenecks, and provide insights for future improvements.

\end{itemize}

\section{Related Works}

\subsection{TQA Datasets}
Early table question answering (TQA) datasets, such as WikiTableQuestions \cite{pasupat2015compositional}, focused on complex queries but provided textual answers only. Spider \cite{spider@emnlp/YuZYYWLMLYRZR18}, while evaluating generalization on complex SQL queries, relies on structured outputs, limiting its open-domain applicability. These datasets are primarily domain-specific.

Recent advancements extend TQA to open-domain scenarios. NQ-TABLES \cite{herzig2021open} introduces candidate table retrieval, and Open-WikiTable \cite{openwikitable@acl/KweonKCJC23} enhances this with metadata. RETQA \cite{RETQA@aaai2025} explores open-domain TQA in real estate, leveraging spoken language understanding (SLU) annotations \cite{xing2025dxa,DBLP:conf/aaai/0001CZW0CL25} for improved retrieval. FutureX \cite{futureX2024} addresses future-oriented QA, focusing on qualitative reasoning rather than structured data. In contrast, ODTQA-FoRe targets precise time-series forecasting over structured tables, a largely overlooked area in open-domain TQA.

\subsection{TQA Methods}
Recent approaches leverage large language models (LLMs) for TQA tasks \cite{TqaSurvey@Fang2024}. OPENTAB \cite{OpenTab@Kong0SSLFRK24} integrates open-domain knowledge for table-based fact-checking, while SLUTQA \cite{RETQA@aaai2025} enhances retrieval and SQL generation accuracy through SLU \cite{SLU2023aaai/ChengY023}. However, these methods mainly focus on historical query answering rather than future predictions.

Independent applications of LLMs for forecasting include TP-BERTa \cite{TP-BERTa}, which discretizes numerical features but does not handle sequential predictions, and LLMTIME \cite{LLMTIME2023Gruver}, which employs pretrained LLMs for time series without TQA integration.

Thus, existing methods predominantly address historical queries, non-sequential forecasting, or time series predictions, lacking a unifying framework for forecasting-oriented TQA. Our work fills this gap, representing the first effort in Open-Domain Tabular Question Answering for Future Data Forecasting and Reasoning.




\section{Dataset Construction and Analysis}

\subsection{Data Source}
\label{data source}
The dataset for this study originates from the RETQA dataset \cite{RETQA@aaai2025}, which covers land auctions, real estate project sales, and corporate finance. Our work focuses on the real estate sales data, initially spanning eight major Chinese cities for the year 2022. We significantly expand this by leveraging the original data source (\url{http://www.fangdi.com.cn/}) to include two additional cities, Tianjin and Chengdu, and extend the timeframe to cover January 2022 to December 2024. This results in a comprehensive dataset of 60,183 real estate projects, each with up to 36 months of data.

To prepare the data for our task, we first define a temporal split: December 31, 2023, serves as the reference date, with data from 2022--2023 designated as historical observations and data from 2024 treated as future ground truth, which is strictly withheld during model inference. Given the inherent sparsity in real estate records, we then apply stringent filtering criteria, retaining only projects with at least six months of sales data in the historical period (2022--2023) and at least one record in the forecast period (2024). After applying these criteria, we obtain a refined dataset of 11,149 projects for a subsequent QA pair generation.

To prevent data leakage and ensure a fair evaluation, we first partition the 11,149 filtered projects into training, validation, and test subsets. This project-level split is performed using a 6:2:2 ratio, resulting in 6,562 projects for the training set, 2,278 for the validation set, and 2,309 for the test set. This partitioning guarantees that there is no overlap of projects between the sets. The QA pairs for each split are subsequently generated exclusively from their corresponding project subsets.

The data corresponding to all 11,149 refined projects is then used to construct the database environments. Since the original RETQA data source is in Chinese, this process primarily involves data cleaning and localization, where all Chinese column headers, city names, district names, and project names are translated into English. The protocols for translation and schema standardization are detailed in Appendix \ref{sec:Appendix-Database-Translation}. After this unified preparation, the dataset is split by time. The historical data (2022--2023) is aggregated by city, district, and year into 288 tables. These tables are then imported into a PostgreSQL \cite{postGRE} database to serve as the agent's queryable knowledge base, each renamed following the format \texttt{project\_price\_table\_\{district\}\_\{city\}\_\{y}-
\texttt{ear\}} (e.g., \texttt{project\_price\_table\_zengcheng\_d}-
\texttt{istrict\_guangzhou\_2022}). Separately, the future data (2024) is organized into 144 tables using the same aggregation and naming convention and stored in an isolated database named ``future prices'' to facilitate the automated execution of ground truth SQL queries during evaluation (see Appendix \ref{sec:Appendix-Template-Filling} for details). Note that throughout these databases, tables do not distinguish between projects from the training, validation, or test sets, as the agent cannot access the full database contents and can only interact with it through a text-to-SQL interface.

\subsection{Template Design}
\label{Template}
To support both time series forecasting and forecast-based reasoning tasks, we design 26 distinct sets of seed templates, comprising 7 templates specifically for forecasting tasks and 19 for reasoning tasks. Each set includes a natural language question template, a historical data SQL template, and a target (future) data SQL template. These templates primarily incorporate four variables: city, district, project name, and time. City, district, and project name variables are populated by randomly sampling from the database, while the time variable is determined based on actual data availability for the respective projects in 2024, ensuring alignment with months that have valid price records.

The forecasting templates generate queries requiring predictions of specific statistical indicators (e.g., price, averages, or extreme values) over predefined periods (see Figure \ref{fig:Query_type}, Q2). Conversely, reasoning templates address more complex analytical tasks, including comparisons of future price trends among multiple projects or identification of projects satisfying specific criteria (see Figure \ref{fig:Query_type}, Q3).

Prior to large-scale generation, we initially create ten QA pairs per template set, totaling 260 pairs. These pairs undergo rigorous manual review to ensure grammatical fluency, semantic clarity, correctness, and efficiency of SQL queries, as well as accuracy and conciseness of results. Any identified issues prompt iterative revisions until all QA pairs pass validation. Subsequently, large-scale automated QA pair generation proceeds. For more details on the types of templates, please refer to Appendix \ref{sec:Appendix-Template-Filling}.

\begin{figure*}[tbp]
 \centering
 \includegraphics[width=0.92\textwidth]{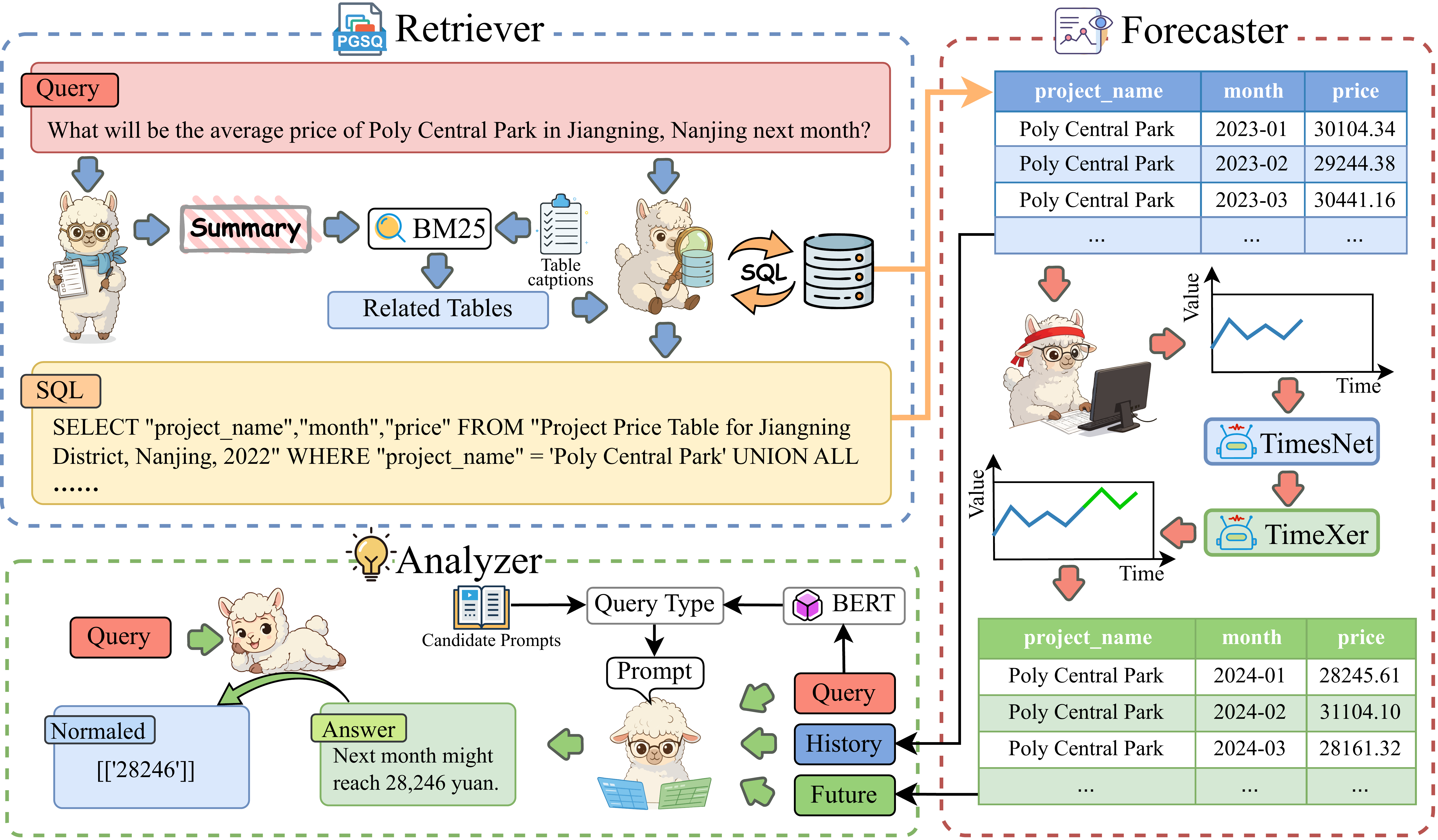}
 \centering
 \caption{General framework of TimeFore.}
 \label{fig:TimeFore_workflow}
\end{figure*}

\subsection{QA Pair Generation}
The final ODTQA-FoRe dataset includes, for each QA instance: a natural language question, either a natural language answer (for reasoning queries) or a numerical answer (for forecasting queries), an SQL query to retrieve historical data, and an SQL query for retrieving future data labels. Historical data query results provide necessary context, while future data query results serve as evaluation labels. Answers are objective, uniquely determined by their corresponding questions. For numerical prediction, we provide only directly relevant values; for reasoning, only the direct conclusion. Multiple values or conclusions, if applicable, are presented as structured lists. Examples and detailed answer formats are provided in Appendix \ref{sec:Appendix-Template-Filling}. All four components (question, answer, historical data SQL, and future data SQL) are generated simultaneously, ensuring precise alignment.


As detailed in Section \ref{Template}, each template utilizes four primary variables: city, district, project name, and time. During QA generation, cities and districts are randomly sampled first, followed by selecting project names from their corresponding dataset splits (training, validation, or test).

To handle temporal references, we create 98 natural language expressions denoting future periods (e.g., ``next month,'' ``next year,'' ``next quarter,'' ``first half of the year''), each mapped to precise month lists (e.g., ``next quarter'' corresponds to [January 2024, February 2024, March 2024]). For each project, we identify all months in 2024 with available records, selecting appropriate expressions fully covered by the data. For instance, ``next quarter'' is used only if data for January to March 2024 is fully available. Subsequently, one eligible expression is randomly selected for template filling. Historical data SQL queries cover all months from 2022–2023, while future data SQL queries explicitly include months in 2024 with available records.

Using this template-driven approach, we initially generate 36,400 QA pairs. After removing duplicates, invalid queries, and empty results, we finalize 28,507 unique QA pairs, divided according to project assignments into 16,944 training pairs, 5,742 validation pairs, and 5,821 testing pairs. Additionally, following previous practices \cite{RETQA@aaai2025}, we employ LLMs to rewrite original questions enhancing dataset diversity and realism. Comprehensive details on the rewriting procedure and dataset statistics are provided in Appendix \ref{sec:Appendix-LLM-based Query Rewriting} and \ref{sec:Appendix-Dataset Statistics}.

\section{Method}

\subsection{Overview}

In this section, we present TimeFore, a collaborative framework designed for Open-Domain Tabular Question Answering involving Future Data Forecasting and Reasoning (ODTQA-FoRe). As illustrated in Figure \ref{fig:TimeFore_workflow}, TimeFore decomposes the task across three specialized agents. The Retriever handles data acquisition by first identifying relevant data tables and then generating a SQL query to extract historical information. Subsequently, the Forecaster leverages function-calling to invoke external, specialized models for accurate time-series prediction. Finally, the Analyzer synthesizes the retrieved data with the generated forecast to construct a precise and consistently formatted final answer.

\subsection{Retriever}
In open-domain scenarios, the performance of accurately retrieving relevant data from a vast repository based on a user's query is paramount, as this step directly governs the quality of all downstream tasks. To address this, the Retriever agent executes a two-stage process. The process commences with table retrieval, where the agent leverages an LLM's summarization capability by employing few-shot prompting with five examples to convert the user query into a concise, canonical text in the style of a table caption. This text is first used to attempt a direct match with existing captions in the database. Should this attempt fail, the BM25 algorithm \cite{robertson1994some} is then employed to retrieve the most semantically relevant table.

Upon securing the target table, the agent proceeds to generate the appropriate SQL query. This is again accomplished through in-context learning, guided by a system prompt containing five distinct exemplars. Each exemplar consists of a natural language question, its corresponding ground-truth table caption, and the target SQL query. To ensure robustness, these examples are sampled to cover diverse query types (e.g., single and multi-item). The LLM is prompted with the user's original question and the identified table caption to generate an initial SQL query. To ensure robustness and address potential syntax or logic errors, particularly in complex scenarios, the Retriever agent employs an execution-feedback loop, utilizing the \textit{sqlQueryTool} to execute the SQL within the database. If execution fails, the agent refines the SQL based on the error feedback. This loop continues until success is achieved or a maximum of 25 iterations is reached, ensuring reliable data retrieval. A detailed illustration of the prompt design is available in Appendix \ref{sec:Appendix-all_prompts}.


\subsection{Forecaster}


Upon receiving the SQL results, the Forecaster interprets the serialized triples formatted as [project name, year–month, price] and converts them into a numerical list for the forecasting tools \cite{DBLP:conf/acl/LiangXLWCCZ25}. This historical input, limited to a 24-month window, remains compact enough to fit within the LLM's context limits, thus avoiding token overflow issues and ensuring smooth data integration for forecasting.

While LLMs possess extensive general knowledge, their native forecasting capabilities for time-series tasks often lag behind specialized models without task-specific fine-tuning \cite{LLMforTS@Dnips2023,mingtian@nips/TanMGAH24}. The Forecaster agent is designed to bridge this gap. Consistent with the TimeFore framework's strategy, it leverages the LLM's function-calling capability to orchestrate specialized, high-performance time-series models, thereby addressing the LLM's inherent predictive weaknesses.

The process begins once the Forecaster receives the historical data sequences from the Retriever. Its core action is to invoke a predefined function, the imputationThenPredictionTool. This function encapsulates a sophisticated forecasting pipeline: it first validates and cleans the input time series, then employs the TimesNet model \cite{TimesNet2023ICLR} for robust data imputation. Based on this complete and imputed sequence, the TimeXer model \cite{wang2024timexer} is then used to generate a price trend forecast for the subsequent 12 months. The tool concludes by returning the results as a structured three-tuple, containing the project name, year-month, and predicted price.

This architecture creates a powerful synergy: the LLM excels at orchestrating the workflow and understanding the task context, while the specialized models handle the precise, mathematical heavy-lifting of forecasting. This division of labor significantly enhances the overall forecasting accuracy and reliability.

For operational simplicity, the Forecaster is configured to generate a full 12-month forecast for the year 2024. This complete forecast is then passed to the Analyzer agent, which is responsible for reasoning upon these results to extract and deliver the specific answers pertinent to the user's original query.

\subsection{Analyzer}

The Analyzer is the final-stage agent in the TimeFore framework, responsible for synthesizing the inputs from the preceding agents into a precise, standardized, and user-centric final answer. Its primary function is to correctly interpret the user's ultimate goal and apply the appropriate reasoning strategy. To this end, the Analyzer first employs a BERT-based classifier to categorize the user's query as either direct time-series forecasting (e.g., "What will the price be in June 2024?") or forecast-based reasoning (e.g., "Will the price in June exceed the price in January?"). Based on this classification, a distinct and tailored prompt, enriched with five illustrative examples via in-context learning, is selected to guide the LLM's response generation process.

To facilitate LLM numerical reasoning, historical and forecast data are formatted as key-value pairs, each entry including a date and its corresponding average price (e.g., ``[[August 2022, 10,890], [December 2022, 11,585]]''). If historical records for specific months are unavailable, these months are simply omitted from the provided data.

Despite its effectiveness in handling forecast-based reasoning queries, LLMs often provide verbose explanations rather than succinct numerical predictions for straightforward forecasting questions \cite{DBLP:journals/corr/abs-2503-09567}. To mitigate this, we introduce an LLM-driven numerical extraction module, which employs in-context learning with five examples, explicitly prompting the model to return only the essential numerical results. All detailed prompts utilized within the analysis agent are comprehensively documented in Appendix \ref{sec:Appendix-all_prompts}.

\section{Experiments}
\begin{table}[htbp]
\begin{tabular}{c|ccc}
\hline
Model       & MSE               & MAE              & MRE             \\ \hline
TimesNet    & 2.77E+07          & 3103.52          & 0.1254          \\
TimeMixer   & 2.78E+07          & 3108.29          & 0.1255          \\
TimeXer     & \textbf{2.50E+07} & \textbf{2989.55} & \textbf{0.1209} \\
WPMixer     & 2.75E+07          & 3097.81          & 0.1248          \\
AutoTimes    & 2.93E+07          & 3204.13          & 0.1288          \\
Time-MoE    & 2.95E+07          & 3164.47          & 0.1271          \\
Qwen3 30B   & 6.69E+07          & 4344.02          & 0.1706          \\
GLM 4.5 Air & 7.30E+07          & 4824.57          & 0.1869          \\ \hline
\end{tabular}
\caption{Performance comparison of different models on time series forecasting tasks.}\label{Time_Series_Forecasting}
\end{table}

\begin{table*}[htbp]
\begin{tabular}{cc|lll|llll}
\hline
\multirow{2}{*}{Model} & \multirow{2}{*}{Method} & \multicolumn{3}{c|}{Time-series Forecasting} & \multicolumn{4}{c}{Forecast-based Reasoning} \\ \cline{3-9} 
 &  & \multicolumn{1}{c}{MSE} & \multicolumn{1}{c}{MAE} & \multicolumn{1}{c|}{MRE} & \multicolumn{1}{c}{Acc} & \multicolumn{1}{c}{P} & \multicolumn{1}{c}{R} & \multicolumn{1}{c}{F1} \\ \hline
\multirow{2}{*}{Qwen3 30B} & Vanilla & 40385720.95 & 3698.36 & 0.1627 & 12.19 & 29.01 & 21.76 & 24.87 \\
 & TimeFore & \textbf{31572410.36} & \textbf{2788.20} & \textbf{0.1326} & \textbf{31.59} & \textbf{61.78} & \textbf{58.80} & \textbf{60.25} \\ \hline
\multirow{2}{*}{Qwen3 Next 80B} & Vanilla & 30942598.21 & 3406.43 & 0.1586 & 24.45 & 48.40 & 44.96 & 46.62 \\
 & TimeFore & \textbf{22442845.96} & \textbf{2588.87} & \textbf{0.1181} & \textbf{36.31} & \textbf{55.86} & \textbf{61.21} & \textbf{58.41} \\ \hline
\multirow{2}{*}{GPT OSS 20B} & Vanilla & 105115117.20 & 4394.56 & 0.1838 & 21.52 & 46.79 & 41.98 & 44.25 \\
 & TimeFore & \textbf{29757828.58} & \textbf{2887.44} & \textbf{0.1280} & \textbf{27.80} & \textbf{53.61} & \textbf{45.31} & \textbf{49.11} \\ \hline
\multirow{2}{*}{GPT OSS 120B} & Vanilla & 47324931.58 & 3786.48 & 0.1683 & 21.60 & 44.44 & 38.16 & 41.06 \\
 & TimeFore & \textbf{18493634.83} & \textbf{2501.18} & \textbf{0.1151} & \textbf{31.37} & \textbf{57.35} & \textbf{47.32} & \textbf{51.86} \\ \hline
\multirow{2}{*}{GLM4.5 Air} & Vanilla & 139922250.13 & 3324.41 & 0.1415 & 23.59 & 49.82 & 47.67 & 48.72 \\
 & TimeFore & \textbf{90865440.59} & \textbf{2709.25} & \textbf{0.1172} & \textbf{35.46} & \textbf{63.73} & \textbf{58.93} & \textbf{61.24} \\ \hline
\end{tabular}
\caption{Overall performance on the ODTQA-FoRe dataset.}\label{Overall_performance}
\end{table*}

\subsection{Experimental Settings and Baselines}

ODTQA-FoRe consists of two subtasks: time-series forecasting and forecast-based reasoning. We evaluate time-series forecasting using Mean Squared Error (MSE), Mean Absolute Error (MAE), and Mean Relative Error (MRE). For forecast-based reasoning, we use Accuracy, Precision (P), Recall (R), and F1-score (F1). To account for questions with multiple items, we compute F1 at the item level for partial credit, whereas Accuracy requires an exact match of all items within a single question.

\noindent\textbf{Data Preprocessing and Model Selection}. Real estate transaction data is often sparse, with information gaps for months without sales. To address this, we first perform data imputation. Our comparison of three leading imputation models, i.e., TimesNet \cite{TimesNet2023ICLR}, i-Transformer \cite{DBLP:conf/iclr/LiuHZWWML24}, and TimeMixer \cite{TimeMixerICLR2024}, shows that TimesNet achieves superior performance, demonstrating the strongest imputation capability (see Appendix \ref{sec:Appendix-Time series imputation} for detailed results).

For the core time-series forecasting task, we evaluate eight models to select the optimal forecaster for our TimeFore framework. These include four lightweight models (TimesNet, TimeMixer, TimeXer \cite{wang2024timexer}, WPMixer \cite{WPMixerAAAI2025}), two LLM-based time-series models (AutoTimes \cite{autoTimesNIPS2024}, Time-MoE \cite{Time-MoE-ICLR2025}), and two general-purpose LLMs (Qwen3 30B, GLM 4.5 Air). The general-purpose LLMs perform forecasting via in-context learning (ICL) on the original data, while the six specialized models are trained on data imputed by TimesNet.

As Table \ref{Time_Series_Forecasting} shows, the specialized time-series models significantly outperform the general-purpose LLMs, which confirms the latter's relative weakness in direct prediction tasks. Among all candidates, TimeXer achieves the best performance and is therefore selected as the forecasting component within the TimeFore framework.

\noindent\textbf{Baselines}.
To validate the effectiveness of TimeFore, we compare it against five strong baseline models: two from the Qwen3 series \cite{yang2025qwen3} (Qwen3 30B A3B, and Qwen3 Next 80B A3B Thinking), two from the GPT-OSS series \cite{openai2025gptoss120bgptoss20bmodel} (GPT-OSS 20B and GPT-OSS 120B), and GLM 4.5 Air  \cite{GLM45report}.

Further details on experimental setups—including retrieval agent performance, training data specifics, and other implementation details—are provided in Appendix \ref{sec:Appendix-Supplementary experiments} due to space constraints.

\subsection{Result and Analysis}

\noindent\textbf{Main Results} 


As  ODTQA-FoRe is a newly proposed benchmark without directly comparable baselines, we introduce a \textit{Vanilla} baseline, which first uses the BM25 algorithm to retrieve relevant table captions, then constructs a five-shot prompt for in-context learning (ICL), and finally relies entirely on the LLM's intrinsic capabilities to predict future data.

As presented in Table \ref{Overall_performance}, the experimental results demonstrate that our TimeFore framework significantly outperforms the vanilla baseline across all models. This outcome highlights the inherent limitations of general-purpose LLMs when applied directly to specialized time-series forecasting tasks.

These results collectively demonstrate that delegating forecasting to a specialized time-series model, rather than relying on the LLM itself, not only enhances prediction accuracy but also significantly improves performance on subsequent reasoning tasks. This finding offers a key insight for the future development of LLM-based predictive and reasoning systems: a hybrid approach that leverages specialized tools for their respective strengths yields a more robust and accurate overall framework.

\subsection{Ablation Study}

\begin{figure}[t]
 \centering
 \includegraphics[width=\columnwidth]{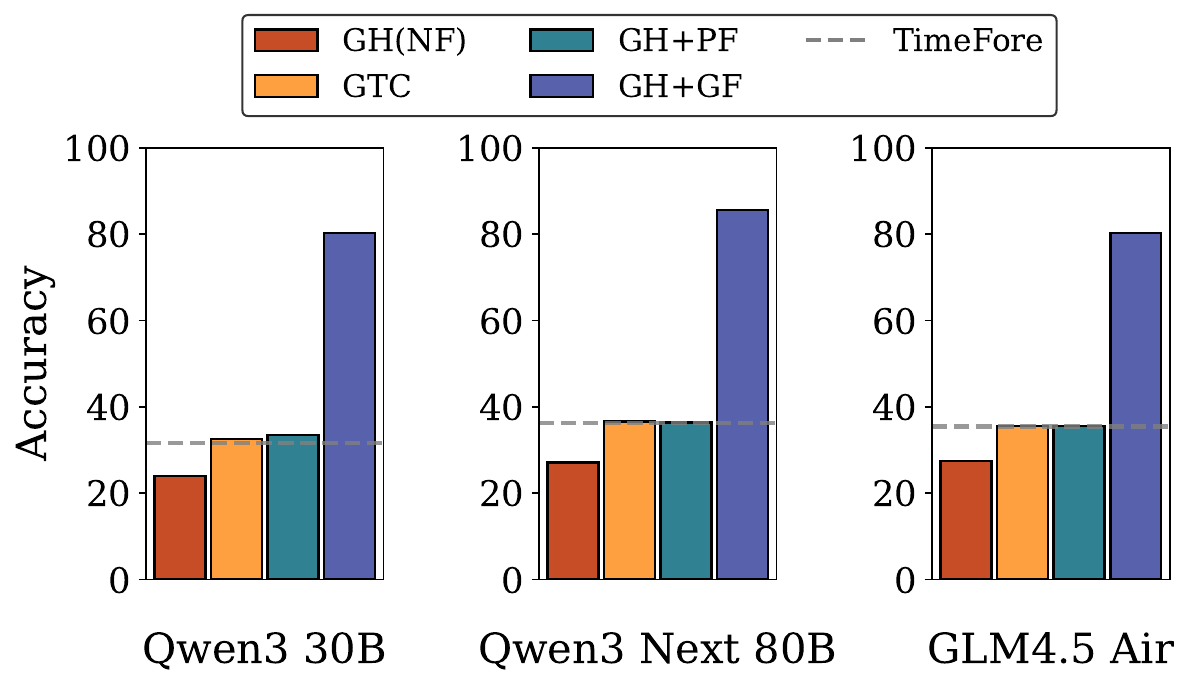}
 \centering
 \caption{Ablation study on the TimeFore framework in forecast-based reasoning tasks.} 
 \label{fig:TSF_Ablation_Study}
\end{figure}

To identify performance bottlenecks and quantify the contribution of each component within our multi-step TimeFore pipeline, we conduct a comprehensive ablation study. For the forecast-based reasoning task, we systematically substitute individual model-generated outputs with their ground-truth counterparts and observe the resulting performance gains. This allows us to isolate the impact of table retrieval, data retrieval (SQL generation), and time-series forecasting.

The study is designed around four specific configurations. The first, denoted as GH (NF) for Golden History (No Future), serves as a baseline where the LLM receives only ground-truth historical data and no future data. This measures the model's reasoning ability without any forecasting input. The second configuration, +GTC, provides ground-truth table captions to the Retriever to isolate the impact of the initial table retrieval step, while the rest of the pipeline remains unchanged. The third setup, GH + PF (Golden History + Predicted Future), bypasses the Retriever by providing ground-truth historical data directly but still requires the Forecaster to predict future values, thereby isolating the performance impact of the SQL generation stage. The final configuration, GH + GF (Golden History + Golden Future), establishes the theoretical upper bound of the LLM's reasoning capability by providing the Analyzer with perfect historical and future data, bypassing both the Retriever and the Forecaster.

The results of this study, visualized in Figure \ref{fig:TSF_Ablation_Study}, reveal a clear performance hierarchy. The most significant performance drop occurs in the GH (NF) setting, confirming that reasoning without forecasting is largely ineffective. While providing golden table captions (+GTC) or golden historical data (GH + PF) yields only marginal improvements, the most substantial performance gain by far is achieved in the GH + GF setting when perfect future data is supplied.

This unequivocally demonstrates that the accuracy of the time-series forecasting component is the primary bottleneck for the entire framework. Accurate numerical prediction is not just beneficial; it is a prerequisite for enabling the large model to perform reliable downstream reasoning.

More detailed ablation study for Analyzer's numerical extraction module are available in Appendix \ref{sec:Appendix-Ablation Study of the Analyzer}.

\section{Conclusion}
In this paper, we introduce a novel task—Open-Domain Tabular Question Answering for Future Data Forecasting and Reasoning—and present a pioneering dataset constructed using real estate data, addressing critical gaps in current research. To support this task, we introduce TimeFore, a comprehensive benchmark framework capable of effectively retrieving historical data, performing accurate future data predictions, and providing standardized responses. Our extensive experiments establish baseline performance across multiple LLMs and highlight both the task's inherent challenges and opportunities, laying the groundwork for future advancements in this field.


\section*{Limitations}

While our study demonstrates the effectiveness of the proposed approach, certain limitations remain that offer opportunities for future improvement. 

First, regarding the forecasting mechanism, we employ TimesNet for data imputation and TimeXer for prediction. While this pipeline is effective, it has not yet incorporated a broader global context, such as regional macroeconomic statistics, environmental factors, or relevant policy information. In future work, we plan to explore models that can better integrate such exogenous variables, especially in scenarios where external factors drive significant market shifts.

Second, regarding domain coverage, the ODTQA-FoRe dataset currently focuses exclusively on the real estate domain. While the TimeFore framework is designed to be domain-agnostic, validating its generalizability across other vertical domains (e.g., finance, retail, or climate) remains a key direction. Specifically, the adaptability of the forecasting backbone (TimeXer) across diverse data characteristics with varying volatility and periodicity warrants further exploration.

Third, regarding dataset construction, while the inclusion of diverse templates may not fully encapsulate the complexity of all real-world queries, the dataset remains sufficiently comprehensive to evaluate the proposed framework. We adopted a template-based generation approach, a common practice in large-scale QA benchmarks, to ensure the availability of deterministic answers. Furthermore, to enhance linguistic diversity, we employed LLMs to rewrite the queries and conducted a human evaluation for validation. The results demonstrate that the rewritten expressions are significantly closer to natural human language than the original template-based formulations.

\section*{Acknowledgments}

This work is supported in part by the National Natural Science Foundation
of China (NSFC) under Grant 62272050 and the grant of Beijing Normal-
Hong Kong Baptist University sponsored
by Guangdong Provincial Department of Education;  in part by Zhuhai Science-Tech Innovation Bureau under
Grant No. 2320004002772 and the Interdisciplinary Intelligence
Super Computer Center of Beijing Normal University (Zhuhai).

\bibliography{custom}

@inproceedings{TP-BERTa,
  author       = {Jiahuan Yan and
                  Bo Zheng and
                  Hongxia Xu and
                  Yiheng Zhu and
                  Danny Z. Chen and
                  Jimeng Sun and
                  Jian Wu and
                  Jintai Chen},
  title        = {Making Pre-trained Language Models Great on Tabular Prediction},
  booktitle    = {The Twelfth International Conference on Learning Representations,
                  {ICLR} 2024, Vienna, Austria, May 7-11, 2024},
  publisher    = {OpenReview.net},
  year         = {2024},
  url          = {https://openreview.net/forum?id=anzIzGZuLi},
  bibsource    = {dblp computer science bibliography, https://dblp.org}
}

@inproceedings{LLMTIME2023Gruver,
  author       = {Nate Gruver and
                  Marc Finzi and
                  Shikai Qiu and
                  Andrew Gordon Wilson},
  editor       = {Alice Oh and
                  Tristan Naumann and
                  Amir Globerson and
                  Kate Saenko and
                  Moritz Hardt and
                  Sergey Levine},
  title        = {Large Language Models Are Zero-Shot Time Series Forecasters},
  booktitle    = {Advances in Neural Information Processing Systems 36: Annual Conference
                  on Neural Information Processing Systems 2023, NeurIPS 2023, New Orleans,
                  LA, USA, December 10 - 16, 2023},
  year         = {2023},
  url          = {http://papers.nips.cc/paper\_files/paper/2023/hash/3eb7ca52e8207697361b2c0fb3926511-Abstract-Conference.html},
  bibsource    = {dblp computer science bibliography, https://dblp.org}
}

@inproceedings{TimesNet2023ICLR,
  author       = {Haixu Wu and
                  Tengge Hu and
                  Yong Liu and
                  Hang Zhou and
                  Jianmin Wang and
                  Mingsheng Long},
  title        = {TimesNet: Temporal 2D-Variation Modeling for General Time Series Analysis},
  booktitle    = {The Eleventh International Conference on Learning Representations,
                  {ICLR} 2023, Kigali, Rwanda, May 1-5, 2023},
  publisher    = {OpenReview.net},
  year         = {2023},
  url          = {https://openreview.net/forum?id=ju\_Uqw384Oq},
  bibsource    = {dblp computer science bibliography, https://dblp.org}
}

@inproceedings{RETQA@aaai2025,
  author       = {Zhensheng Wang and
                  Wenmian Yang and
                  Kun Zhou and
                  Yiquan Zhang and
                  Weijia Jia},
  editor       = {Toby Walsh and
                  Julie Shah and
                  Zico Kolter},
  title        = {{RETQA:} {A} Large-Scale Open-Domain Tabular Question Answering Dataset
                  for Real Estate Sector},
  booktitle    = {AAAI-25, Sponsored by the Association for the Advancement of Artificial
                  Intelligence, February 25 - March 4, 2025, Philadelphia, PA, {USA}},
  pages        = {25452--25460},
  publisher    = {{AAAI} Press},
  year         = {2025},
  url          = {https://doi.org/10.1609/aaai.v39i24.34734},
  doi          = {10.1609/AAAI.V39I24.34734},
  bibsource    = {dblp computer science bibliography, https://dblp.org}
}

@inproceedings{LLMforTS@Dnips2023,
  author       = {Nate Gruver and
                  Marc Finzi and
                  Shikai Qiu and
                  Andrew Gordon Wilson},
  editor       = {Alice Oh and
                  Tristan Naumann and
                  Amir Globerson and
                  Kate Saenko and
                  Moritz Hardt and
                  Sergey Levine},
  title        = {Large Language Models Are Zero-Shot Time Series Forecasters},
  booktitle    = {Advances in Neural Information Processing Systems 36: Annual Conference
                  on Neural Information Processing Systems 2023, NeurIPS 2023, New Orleans,
                  LA, USA, December 10 - 16, 2023},
  year         = {2023},
  url          = {http://papers.nips.cc/paper\_files/paper/2023/hash/3eb7ca52e8207697361b2c0fb3926511-Abstract-Conference.html},
  bibsource    = {dblp computer science bibliography, https://dblp.org}
}

@inproceedings{OpenTab@Kong0SSLFRK24,
  author       = {Kezhi Kong and
                  Jiani Zhang and
                  Zhengyuan Shen and
                  Balasubramaniam Srinivasan and
                  Chuan Lei and
                  Christos Faloutsos and
                  Huzefa Rangwala and
                  George Karypis},
  title        = {OpenTab: Advancing Large Language Models as Open-domain Table Reasoners},
  booktitle    = {The Twelfth International Conference on Learning Representations,
                  {ICLR} 2024, Vienna, Austria, May 7-11, 2024},
  publisher    = {OpenReview.net},
  year         = {2024},
  url          = {https://openreview.net/forum?id=Qa0ULgosc9},
  bibsource    = {dblp computer science bibliography, https://dblp.org}
}

@Article{TqaSurvey@Fang2024,
 author = {Xi Fang and Weijie Xu and Fiona Anting Tan and Jiani Zhang and Ziqing Hu and Yanjun (Jane) Qi and Scott Nickleach and Diego Socolinsky and Srinivasan Sengamedu, "SHS" and Christos Faloutsos},
 title = {Large language models (LLMs) on tabular data: Prediction, generation, and understanding — a survey},
 year = {2024},
 url = {https://www.amazon.science/publications/large-language-models-llms-on-tabular-data-prediction-generation-and-understanding-a-survey},
 journal = {Transactions on Machine Learning Research},
}

@inproceedings{mingtian@nips/TanMGAH24,
  author       = {Mingtian Tan and
                  Mike A. Merrill and
                  Vinayak Gupta and
                  Tim Althoff and
                  Tom Hartvigsen},
  editor       = {Amir Globersons and
                  Lester Mackey and
                  Danielle Belgrave and
                  Angela Fan and
                  Ulrich Paquet and
                  Jakub M. Tomczak and
                  Cheng Zhang},
  title        = {Are Language Models Actually Useful for Time Series Forecasting?},
  booktitle    = {Advances in Neural Information Processing Systems 38: Annual Conference
                  on Neural Information Processing Systems 2024, NeurIPS 2024, Vancouver,
                  BC, Canada, December 10 - 15, 2024},
  year         = {2024},
  url          = {http://papers.nips.cc/paper\_files/paper/2024/hash/6ed5bf446f59e2c6646d23058c86424b-Abstract-Conference.html},
  bibsource    = {dblp computer science bibliography, https://dblp.org}
}

@inproceedings{spider@emnlp/YuZYYWLMLYRZR18,
  author       = {Tao Yu and
                  Rui Zhang and
                  Kai Yang and
                  Michihiro Yasunaga and
                  Dongxu Wang and
                  Zifan Li and
                  James Ma and
                  Irene Li and
                  Qingning Yao and
                  Shanelle Roman and
                  Zilin Zhang and
                  Dragomir R. Radev},
  editor       = {Ellen Riloff and
                  David Chiang and
                  Julia Hockenmaier and
                  Jun'ichi Tsujii},
  title        = {Spider: {A} Large-Scale Human-Labeled Dataset for Complex and Cross-Domain
                  Semantic Parsing and Text-to-SQL Task},
  booktitle    = {Proceedings of the 2018 Conference on Empirical Methods in Natural
                  Language Processing, Brussels, Belgium, October 31 - November 4, 2018},
  pages        = {3911--3921},
  publisher    = {Association for Computational Linguistics},
  year         = {2018},
  url          = {https://doi.org/10.18653/v1/d18-1425},
  doi          = {10.18653/V1/D18-1425},
  bibsource    = {dblp computer science bibliography, https://dblp.org}
}

@inproceedings{openwikitable@acl/KweonKCJC23,
  author       = {Sunjun Kweon and
                  Yeonsu Kwon and
                  Seonhee Cho and
                  Yohan Jo and
                  Edward Choi},
  editor       = {Anna Rogers and
                  Jordan L. Boyd{-}Graber and
                  Naoaki Okazaki},
  title        = {Open-WikiTable : Dataset for Open Domain Question Answering with Complex
                  Reasoning over Table},
  booktitle    = {Findings of the Association for Computational Linguistics: {ACL} 2023,
                  Toronto, Canada, July 9-14, 2023},
  pages        = {8285--8297},
  publisher    = {Association for Computational Linguistics},
  year         = {2023},
  url          = {https://doi.org/10.18653/v1/2023.findings-acl.526},
  doi          = {10.18653/V1/2023.FINDINGS-ACL.526},
  bibsource    = {dblp computer science bibliography, https://dblp.org}
}

@inproceedings{pasupat2015compositional,
  title={Compositional Semantic Parsing on Semi-Structured Tables},
  author={Pasupat, Panupong and Liang, Percy},
  booktitle={Proceedings of the 53rd Annual Meeting of the Association for Computational Linguistics and the 7th International Joint Conference on Natural Language Processing (Volume 1: Long Papers)},
  pages={1470--1480},
  year={2015}
}

@inproceedings{ChenFWB2017ACL,
  author       = {Danqi Chen and
                  Adam Fisch and
                  Jason Weston and
                  Antoine Bordes},
  editor       = {Regina Barzilay and
                  Min{-}Yen Kan},
  title        = {Reading Wikipedia to Answer Open-Domain Questions},
  booktitle    = {Proceedings of the 55th Annual Meeting of the Association for Computational
                  Linguistics, {ACL} 2017, Vancouver, Canada, July 30 - August 4, Volume
                  1: Long Papers},
  pages        = {1870--1879},
  publisher    = {Association for Computational Linguistics},
  year         = {2017},
  url          = {https://doi.org/10.18653/v1/P17-1171},
  doi          = {10.18653/V1/P17-1171},
  bibsource    = {dblp computer science bibliography, https://dblp.org}
}

@inproceedings{herzig2021open,
  author       = {Jonathan Herzig and
                  Thomas M{\"{u}}ller and
                  Syrine Krichene and
                  Julian Martin Eisenschlos},
  editor       = {Kristina Toutanova and
                  Anna Rumshisky and
                  Luke Zettlemoyer and
                  Dilek Hakkani{-}T{\"{u}}r and
                  Iz Beltagy and
                  Steven Bethard and
                  Ryan Cotterell and
                  Tanmoy Chakraborty and
                  Yichao Zhou},
  title        = {Open Domain Question Answering over Tables via Dense Retrieval},
  booktitle    = {Proceedings of the 2021 Conference of the North American Chapter of
                  the Association for Computational Linguistics: Human Language Technologies,
                  {NAACL-HLT} 2021, Online, June 6-11, 2021},
  pages        = {512--519},
  publisher    = {Association for Computational Linguistics},
  year         = {2021},
  url          = {https://doi.org/10.18653/v1/2021.naacl-main.43},
  doi          = {10.18653/V1/2021.NAACL-MAIN.43},
  bibsource    = {dblp computer science bibliography, https://dblp.org}
}

@inproceedings{robertson1994some,
  title={Some simple effective approximations to the 2-poisson model for probabilistic weighted retrieval},
  author={Robertson, Stephen E and Walker, Steve},
  booktitle={SIGIR’94: Proceedings of the Seventeenth Annual International ACM-SIGIR Conference on Research and Development in Information Retrieval, organised by Dublin City University},
  pages={232--241},
  year={1994},
  organization={Springer}
}

@inproceedings{SLU2023aaai/ChengY023,
  author       = {Lizhi Cheng and
                  Wenmian Yang and
                  Weijia Jia},
  editor       = {Brian Williams and
                  Yiling Chen and
                  Jennifer Neville},
  title        = {A Scope Sensitive and Result Attentive Model for Multi-Intent Spoken
                  Language Understanding},
  booktitle    = {Thirty-Seventh {AAAI} Conference on Artificial Intelligence, {AAAI}
                  2023, Thirty-Fifth Conference on Innovative Applications of Artificial
                  Intelligence, {IAAI} 2023, Thirteenth Symposium on Educational Advances
                  in Artificial Intelligence, {EAAI} 2023, Washington, DC, USA, February
                  7-14, 2023},
  pages        = {12691--12699},
  publisher    = {{AAAI} Press},
  year         = {2023},
  url          = {https://doi.org/10.1609/aaai.v37i11.26493},
  doi          = {10.1609/AAAI.V37I11.26493},
  bibsource    = {dblp computer science bibliography, https://dblp.org}
}

@inproceedings{openQA2023nips,
  author       = {Cunxiang Wang and
                  Sirui Cheng and
                  Qipeng Guo and
                  Yuanhao Yue and
                  Bowen Ding and
                  Zhikun Xu and
                  Yidong Wang and
                  Xiangkun Hu and
                  Zheng Zhang and
                  Yue Zhang},
  editor       = {Alice Oh and
                  Tristan Naumann and
                  Amir Globerson and
                  Kate Saenko and
                  Moritz Hardt and
                  Sergey Levine},
  title        = {Evaluating Open-QA Evaluation},
  booktitle    = {Advances in Neural Information Processing Systems 36: Annual Conference
                  on Neural Information Processing Systems 2023, NeurIPS 2023, New Orleans,
                  LA, USA, December 10 - 16, 2023},
  year         = {2023},
  url          = {http://papers.nips.cc/paper\_files/paper/2023/hash/f323d594aa5d2c68154433a131c07959-Abstract-Datasets\_and\_Benchmarks.html},
  bibsource    = {dblp computer science bibliography, https://dblp.org}
}

@inproceedings{TabLaP@aaai2025Wang,
  author       = {Yuxiang Wang and
                  Jianzhong Qi and
                  Junhao Gan},
  editor       = {Toby Walsh and
                  Julie Shah and
                  Zico Kolter},
  title        = {Accurate and Regret-Aware Numerical Problem Solver for Tabular Question
                  Answering},
  booktitle    = {AAAI-25, Sponsored by the Association for the Advancement of Artificial
                  Intelligence, February 25 - March 4, 2025, Philadelphia, PA, {USA}},
  pages        = {12775--12783},
  publisher    = {{AAAI} Press},
  year         = {2025},
  url          = {https://doi.org/10.1609/aaai.v39i12.33393},
  doi          = {10.1609/AAAI.V39I12.33393},
  bibsource    = {dblp computer science bibliography, https://dblp.org}
}

@inproceedings{RAG2020nips,
  author       = {Patrick Lewis and
                  Ethan Perez and
                  Aleksandra Piktus and
                  Fabio Petroni and
                  Vladimir Karpukhin and
                  Naman Goyal and
                  Heinrich K{\"{u}}ttler and
                  Mike Lewis and
                  Wen{-}tau Yih and
                  Tim Rockt{\"{a}}schel and
                  Sebastian Riedel and
                  Douwe Kiela},
  editor       = {Hugo Larochelle and
                  Marc'Aurelio Ranzato and
                  Raia Hadsell and
                  Maria{-}Florina Balcan and
                  Hsuan{-}Tien Lin},
  title        = {Retrieval-Augmented Generation for Knowledge-Intensive {NLP} Tasks},
  booktitle    = {Advances in Neural Information Processing Systems 33: Annual Conference
                  on Neural Information Processing Systems 2020, NeurIPS 2020, December
                  6-12, 2020, virtual},
  year         = {2020},
  url          = {https://proceedings.neurips.cc/paper/2020/hash/6b493230205f780e1bc26945df7481e5-Abstract.html},
  bibsource    = {dblp computer science bibliography, https://dblp.org}
}

@article{DBLP:journals/corr/abs-2503-09567,
  author       = {Qiguang Chen and
                  Libo Qin and
                  Jinhao Liu and
                  Dengyun Peng and
                  Jiannan Guan and
                  Peng Wang and
                  Mengkang Hu and
                  Yuhang Zhou and
                  Te Gao and
                  Wanxiang Che},
  title        = {Towards Reasoning Era: {A} Survey of Long Chain-of-Thought for Reasoning
                  Large Language Models},
  journal      = {CoRR},
  volume       = {abs/2503.09567},
  year         = {2025},
  url          = {https://doi.org/10.48550/arXiv.2503.09567},
  doi          = {10.48550/ARXIV.2503.09567},
  eprinttype    = {arXiv},
  eprint       = {2503.09567},
  bibsource    = {dblp computer science bibliography, https://dblp.org}
}

@incollection{postGRE,
  author       = {Michael Stonebraker and
                  Lawrence A. Rowe and
                  Michael Hirohama},
  editor       = {Michael L. Brodie},
  title        = {The implementation of {POSTGRES}},
  booktitle    = {Making Databases Work: the Pragmatic Wisdom of Michael Stonebraker},
  series       = {{ACM} Books},
  volume       = {22},
  pages        = {519--559},
  publisher    = {{ACM} / Morgan {\&} Claypool},
  year         = {2019},
  url          = {https://doi.org/10.1145/3226595.3226639},
  doi          = {10.1145/3226595.3226639},
  bibsource    = {dblp computer science bibliography, https://dblp.org}
}

@article{wang2024timexer,
  title={Timexer: Empowering transformers for time series forecasting with exogenous variables},
  author={Wang, Yuxuan and Wu, Haixu and Dong, Jiaxiang and Qin, Guo and Zhang, Haoran and Liu, Yong and Qiu, Yunzhong and Wang, Jianmin and Long, Mingsheng},
  journal={Advances in Neural Information Processing Systems},
  volume={37},
  pages={469--498},
  year={2024}
}

@article{yang2025qwen3,
  title={Qwen3 technical report},
  author={Yang, An and Li, Anfeng and Yang, Baosong and Zhang, Beichen and Hui, Binyuan and Zheng, Bo and Yu, Bowen and Gao, Chang and Huang, Chengen and Lv, Chenxu and others},
  journal={arXiv preprint arXiv:2505.09388},
  year={2025}
}

@misc{openai2025gptoss120bgptoss20bmodel,
      title={gpt-oss-120b \& gpt-oss-20b Model Card}, 
      author={Agarwal, Sandhini and Ahmad, Lama and Ai, Jason and Altman, Sam and Applebaum, Andy and Arbus, Edwin and Arora, Rahul K and Bai, Yu and Baker, Bowen and Bao, Haiming and others},
      year={2025},
      eprint={2508.10925},
      archivePrefix={arXiv},
      primaryClass={cs.CL},
      url={https://arxiv.org/abs/2508.10925}, 
}

@inproceedings{DBLP:conf/aaai/0001CZW0CL25,
  author       = {Libo Qin and
                  Qiguang Chen and
                  Jingxuan Zhou and
                  Jin Wang and
                  Hao Fei and
                  Wanxiang Che and
                  Min Li},
  editor       = {Toby Walsh and
                  Julie Shah and
                  Zico Kolter},
  title        = {Divide-Solve-Combine: An Interpretable and Accurate Prompting Framework
                  for Zero-shot Multi-Intent Detection},
  booktitle    = {AAAI-25, Sponsored by the Association for the Advancement of Artificial
                  Intelligence, February 25 - March 4, 2025, Philadelphia, PA, {USA}},
  pages        = {25038--25046},
  publisher    = {{AAAI} Press},
  year         = {2025},
  url          = {https://doi.org/10.1609/aaai.v39i23.34688},
  doi          = {10.1609/AAAI.V39I23.34688},
  bibsource    = {dblp computer science bibliography, https://dblp.org}
}

@article{DBLP:journals/fcsc/ZhangWDZC25,
  author       = {Xuanliang Zhang and
                  Dingzirui Wang and
                  Longxu Dou and
                  Qingfu Zhu and
                  Wanxiang Che},
  title        = {A survey of table reasoning with large language models},
  journal      = {Frontiers Comput. Sci.},
  volume       = {19},
  number       = {9},
  pages        = {199348},
  year         = {2025},
  url          = {https://doi.org/10.1007/s11704-024-40330-z},
  doi          = {10.1007/S11704-024-40330-Z},
  bibsource    = {dblp computer science bibliography, https://dblp.org}
}

@article{xing2025dxa,
  title={DXA-Net: Dual-task Cross-lingual Alignment Network for Zero-shot Cross-lingual Spoken Language Understanding},
  author={Xing, Bowen and Qin, Libo and Zhu, Zhihong and Yu, Zhou and Tsang, Ivor W},
  journal={IEEE Transactions on Pattern Analysis and Machine Intelligence},
  year={2025},
  publisher={IEEE}
}

@article{GLM45report,
  author       = {Aohan Zeng and
                  Xin Lv and
                  Qinkai Zheng and
                  Zhenyu Hou and
                  Bin Chen and
                  Chengxing Xie and
                  Cunxiang Wang and
                  Da Yin and
                  Hao Zeng and
                  Jiajie Zhang and
                  Kedong Wang and
                  Lucen Zhong and
                  Mingdao Liu and
                  Rui Lu and
                  Shulin Cao and
                  Xiaohan Zhang and
                  Xuancheng Huang and
                  Yao Wei and
                  Yean Cheng and
                  Yifan An and
                  Yilin Niu and
                  Yuanhao Wen and
                  Yushi Bai and
                  Zhengxiao Du and
                  Zihan Wang and
                  Zilin Zhu and
                  Bohan Zhang and
                  Bosi Wen and
                  Bowen Wu and
                  Bowen Xu and
                  Can Huang and
                  Casey Zhao and
                  Changpeng Cai and
                  Chao Yu and
                  Chen Li and
                  Chendi Ge and
                  Chenghua Huang and
                  Chenhui Zhang and
                  Chenxi Xu and
                  Chenzheng Zhu and
                  Chuang Li and
                  Congfeng Yin and
                  Daoyan Lin and
                  Dayong Yang and
                  Dazhi Jiang and
                  Ding Ai and
                  Erle Zhu and
                  Fei Wang and
                  Gengzheng Pan and
                  Guo Wang and
                  Hailong Sun and
                  Haitao Li and
                  Haiyang Li and
                  Haiyi Hu and
                  Hanyu Zhang and
                  Hao Peng and
                  Hao Tai and
                  Haoke Zhang and
                  Haoran Wang and
                  Haoyu Yang and
                  He Liu and
                  He Zhao and
                  Hongwei Liu and
                  Hongxi Yan and
                  Huan Liu and
                  Huilong Chen and
                  Ji Li and
                  Jiajing Zhao and
                  Jiamin Ren and
                  Jian Jiao and
                  Jiani Zhao and
                  Jianyang Yan and
                  Jiaqi Wang and
                  Jiayi Gui and
                  Jiayue Zhao and
                  Jie Liu and
                  Jijie Li and
                  Jing Li and
                  Jing Lu and
                  Jingsen Wang and
                  Jingwei Yuan and
                  Jingxuan Li and
                  Jingzhao Du and
                  Jinhua Du and
                  Jinxin Liu and
                  Junkai Zhi and
                  Junli Gao and
                  Ke Wang and
                  Lekang Yang and
                  Liang Xu and
                  Lin Fan and
                  Lindong Wu and
                  Lintao Ding and
                  Lu Wang and
                  Man Zhang and
                  Minghao Li and
                  Minghuan Xu and
                  Mingming Zhao and
                  Mingshu Zhai},
  title        = {{GLM-4.5:} Agentic, Reasoning, and Coding {(ARC)} Foundation Models},
  journal      = {CoRR},
  volume       = {abs/2508.06471},
  year         = {2025},
  url          = {https://doi.org/10.48550/arXiv.2508.06471},
  doi          = {10.48550/ARXIV.2508.06471},
  eprinttype    = {arXiv},
  eprint       = {2508.06471},
  bibsource    = {dblp computer science bibliography, https://dblp.org}
}

@inproceedings{TimeMixerICLR2024,
  author       = {Shiyu Wang and
                  Haixu Wu and
                  Xiaoming Shi and
                  Tengge Hu and
                  Huakun Luo and
                  Lintao Ma and
                  James Y. Zhang and
                  Jun Zhou},
  title        = {TimeMixer: Decomposable Multiscale Mixing for Time Series Forecasting},
  booktitle    = {The Twelfth International Conference on Learning Representations,
                  {ICLR} 2024, Vienna, Austria, May 7-11, 2024},
  publisher    = {OpenReview.net},
  year         = {2024},
  url          = {https://openreview.net/forum?id=7oLshfEIC2},
  bibsource    = {dblp computer science bibliography, https://dblp.org}
}

@inproceedings{WPMixerAAAI2025,
  author       = {Md Mahmuddun Nabi Murad and
                  Mehmet Aktukmak and
                  Yasin Yilmaz},
  editor       = {Toby Walsh and
                  Julie Shah and
                  Zico Kolter},
  title        = {WPMixer: Efficient Multi-Resolution Mixing for Long-Term Time Series
                  Forecasting},
  booktitle    = {AAAI-25, Sponsored by the Association for the Advancement of Artificial
                  Intelligence, February 25 - March 4, 2025, Philadelphia, PA, {USA}},
  pages        = {19581--19588},
  publisher    = {{AAAI} Press},
  year         = {2025},
  url          = {https://doi.org/10.1609/aaai.v39i18.34156},
  doi          = {10.1609/AAAI.V39I18.34156},
  bibsource    = {dblp computer science bibliography, https://dblp.org}
}

@inproceedings{autoTimesNIPS2024,
  author       = {Yong Liu and
                  Guo Qin and
                  Xiangdong Huang and
                  Jianmin Wang and
                  Mingsheng Long},
  editor       = {Amir Globersons and
                  Lester Mackey and
                  Danielle Belgrave and
                  Angela Fan and
                  Ulrich Paquet and
                  Jakub M. Tomczak and
                  Cheng Zhang},
  title        = {AutoTimes: Autoregressive Time Series Forecasters via Large Language
                  Models},
  booktitle    = {Advances in Neural Information Processing Systems 38: Annual Conference
                  on Neural Information Processing Systems 2024, NeurIPS 2024, Vancouver,
                  BC, Canada, December 10 - 15, 2024},
  year         = {2024},
  url          = {http://papers.nips.cc/paper\_files/paper/2024/hash/dcf88cbc8d01ce7309b83d0ebaeb9d29-Abstract-Conference.html},
  bibsource    = {dblp computer science bibliography, https://dblp.org}
}

@inproceedings{Time-MoE-ICLR2025,
  author       = {Xiaoming Shi and
                  Shiyu Wang and
                  Yuqi Nie and
                  Dianqi Li and
                  Zhou Ye and
                  Qingsong Wen and
                  Ming Jin},
  title        = {Time-MoE: Billion-Scale Time Series Foundation Models with Mixture
                  of Experts},
  booktitle    = {The Thirteenth International Conference on Learning Representations,
                  {ICLR} 2025, Singapore, April 24-28, 2025},
  publisher    = {OpenReview.net},
  year         = {2025},
  url          = {https://openreview.net/forum?id=e1wDDFmlVu},
  bibsource    = {dblp computer science bibliography, https://dblp.org}
}

@inproceedings{DBLP:conf/iclr/LiuHZWWML24,
  author       = {Yong Liu and
                  Tengge Hu and
                  Haoran Zhang and
                  Haixu Wu and
                  Shiyu Wang and
                  Lintao Ma and
                  Mingsheng Long},
  title        = {iTransformer: Inverted Transformers Are Effective for Time Series
                  Forecasting},
  booktitle    = {The Twelfth International Conference on Learning Representations,
                  {ICLR} 2024, Vienna, Austria, May 7-11, 2024},
  publisher    = {OpenReview.net},
  year         = {2024},
  url          = {https://openreview.net/forum?id=JePfAI8fah},
  bibsource    = {dblp computer science bibliography, https://dblp.org}
}

@article{futureX2024,
  author       = {Zhiyuan Zeng and
                  Jiashuo Liu and
                  Siyuan Chen and
                  Tianci He and
                  Yali Liao and
                  Jinpeng Wang and
                  Zaiyuan Wang and
                  Yang Yang and
                  Lingyue Yin and
                  Mingren Yin and
                  Zhenwei Zhu and
                  Tianle Cai and
                  Zehui Chen and
                  Jiecao Chen and
                  Yantao Du and
                  Xiang Gao and
                  Jiacheng Guo and
                  Liang Hu and
                  Jianpeng Jiao and
                  Xiangsheng Li and
                  Jingkai Liu and
                  Shuang Ni and
                  Zhoufutu Wen and
                  Ge Zhang and
                  Kaiyuan Zhang and
                  Xin Zhou and
                  Jose Blanchet and
                  Xipeng Qiu and
                  Mengdi Wang and
                  Wenhao Huang},
  title        = {FutureX: An Advanced Live Benchmark for {LLM} Agents in Future Prediction},
  journal      = {CoRR},
  volume       = {abs/2508.11987},
  year         = {2025},
  url          = {https://doi.org/10.48550/arXiv.2508.11987},
  doi          = {10.48550/ARXIV.2508.11987},
  eprinttype    = {arXiv},
  eprint       = {2508.11987},
  bibsource    = {dblp computer science bibliography, https://dblp.org}
}

@inproceedings{DBLP:conf/acl/ChenL023,
  author       = {Ziyang Chen and
                  Jinzhi Liao and
                  Xiang Zhao},
  editor       = {Anna Rogers and
                  Jordan L. Boyd{-}Graber and
                  Naoaki Okazaki},
  title        = {Multi-granularity Temporal Question Answering over Knowledge Graphs},
  booktitle    = {Proceedings of the 61st Annual Meeting of the Association for Computational
                  Linguistics (Volume 1: Long Papers), {ACL} 2023, Toronto, Canada,
                  July 9-14, 2023},
  pages        = {11378--11392},
  publisher    = {Association for Computational Linguistics},
  year         = {2023},
  url          = {https://doi.org/10.18653/v1/2023.acl-long.637},
  doi          = {10.18653/V1/2023.ACL-LONG.637},
  bibsource    = {dblp computer science bibliography, https://dblp.org}
}

@inproceedings{DBLP:conf/nips/ChengYFGYK0L24,
  author       = {An{-}Chieh Cheng and
                  Hongxu Yin and
                  Yang Fu and
                  Qiushan Guo and
                  Ruihan Yang and
                  Jan Kautz and
                  Xiaolong Wang and
                  Sifei Liu},
  editor       = {Amir Globersons and
                  Lester Mackey and
                  Danielle Belgrave and
                  Angela Fan and
                  Ulrich Paquet and
                  Jakub M. Tomczak and
                  Cheng Zhang},
  title        = {SpatialRGPT: Grounded Spatial Reasoning in Vision-Language Models},
  booktitle    = {Advances in Neural Information Processing Systems 38: Annual Conference
                  on Neural Information Processing Systems 2024, NeurIPS 2024, Vancouver,
                  BC, Canada, December 10 - 15, 2024},
  year         = {2024},
  url          = {http://papers.nips.cc/paper\_files/paper/2024/hash/f38cb4cf9a5eaa92b3cfa481832719c6-Abstract-Conference.html},
  bibsource    = {dblp computer science bibliography, https://dblp.org}
}

@inproceedings{DBLP:conf/acl/LiangXLWCCZ25,
  author       = {Xiaobo Liang and
                  Wenjin Xie and
                  Juntao Li and
                  Wanfu Wang and
                  Yibin Chen and
                  Kehai Chen and
                  Min Zhang},
  editor       = {Wanxiang Che and
                  Joyce Nabende and
                  Ekaterina Shutova and
                  Mohammad Taher Pilehvar},
  title        = {Tool learning via Inference-time Scaling and Cycle Verifier},
  booktitle    = {Findings of the Association for Computational Linguistics, {ACL} 2025,
                  Vienna, Austria, July 27 - August 1, 2025},
  series       = {Findings of {ACL}},
  pages        = {24658--24671},
  publisher    = {Association for Computational Linguistics},
  year         = {2025},
  url          = {https://aclanthology.org/2025.findings-acl.1266/},
  bibsource    = {dblp computer science bibliography, https://dblp.org}
}

@inproceedings{chen-etal-2025-llm-based,
    title = "{LLM}-based Translation Inference with Iterative Bilingual Understanding",
    author = "Chen, Andong  and
      Chen, Kehai  and
      Xiang, Yang  and
      Bai, Xuefeng  and
      Yang, Muyun  and
      Feng, Yang  and
      Zhao, Tiejun  and
      Zhang, Min",
    editor = "Che, Wanxiang  and
      Nabende, Joyce  and
      Shutova, Ekaterina  and
      Pilehvar, Mohammad Taher",
    booktitle = "Findings of the Association for Computational Linguistics: ACL 2025",
    month = jul,
    year = "2025",
    address = "Vienna, Austria",
    publisher = "Association for Computational Linguistics",
    url = "https://aclanthology.org/2025.findings-acl.867/",
    doi = "10.18653/v1/2025.findings-acl.867",
    pages = "16886--16902",
    ISBN = "979-8-89176-256-5"
}

@inproceedings{ye-etal-2025-optimizing,
    title = "Optimizing Question Semantic Space for Dynamic Retrieval-Augmented Multi-hop Question Answering",
    author = "Ye, Linhao  and
      Yu, Lang  and
      Lei, Zhikai  and
      Chen, Qin  and
      Zhou, Jie  and
      He, Liang",
    editor = "Che, Wanxiang  and
      Nabende, Joyce  and
      Shutova, Ekaterina  and
      Pilehvar, Mohammad Taher",
    booktitle = "Proceedings of the 63rd Annual Meeting of the Association for Computational Linguistics (Volume 1: Long Papers)",
    month = jul,
    year = "2025",
    address = "Vienna, Austria",
    publisher = "Association for Computational Linguistics",
    url = "https://aclanthology.org/2025.acl-long.871/",
    doi = "10.18653/v1/2025.acl-long.871",
    pages = "17814--17824",
    ISBN = "979-8-89176-251-0"
}

\appendix

\section{Datasets}
\label{sec:appendix-Datasets}

\subsection{Database Translation and Standardization}
\label{sec:Appendix-Database-Translation}

Since the original data source (RETQA) is in Chinese, a rigorous protocol for translation and normalization is essential to ensure compatibility with English-based LLMs and reproducibility of the SQL-based workflows. This process is divided into handling structured schema elements and entities.

\textbf{Schema Normalization.}
First, we standardized the structured architecture elements, including column headers, city names, and district names, by mapping them to unified English terms. This manually curated mapping ensures consistency across different tables. For instance, the column representing project names is standardized as ``project\_name'' in all tables, and district names use standard Pinyin or official English translations. This step is a prerequisite for the Retriever agent to generate valid and executable SQL queries without ambiguity.

\begin{figure*}[hp]
 \centering
 \includegraphics[width=\textwidth]{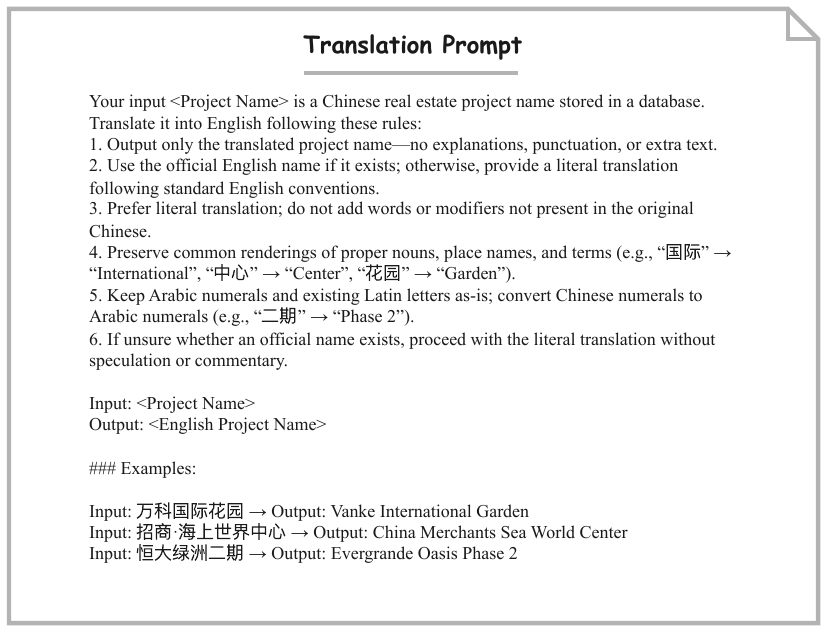}
 \centering
 \caption{Prompt for LLM to translate project names.}
 \label{fig:Translate_prompt}
\end{figure*}

\textbf{Entity Translation.}
Translating real estate project names presents a greater challenge due to semantic variability and the lack of a standard dictionary. To address this while maintaining semantic fidelity and consistency, we employed an LLM-driven translation strategy. As illustrated in Figure \ref{fig:Translate_prompt}, we designed a specialized prompt that constrains the model's output to follow strict translation rules.

\textbf{Quality Assurance.}
To address potential concerns regarding translation errors affecting downstream retrieval or reasoning, we implemented a validation step. Specifically, we conducted a manual review of a random sample of translated project names to verify adherence to the standardized rules. This validation confirmed that the LLM, guided by the prompt, successfully generated consistent English names that align with the database schema, thereby minimizing the risk of retrieval failure due to naming mismatches.

\subsection{Template Filling}
\label{sec:Appendix-Template-Filling}

\begin{figure}[h]
 \centering
 \includegraphics[width=\columnwidth]{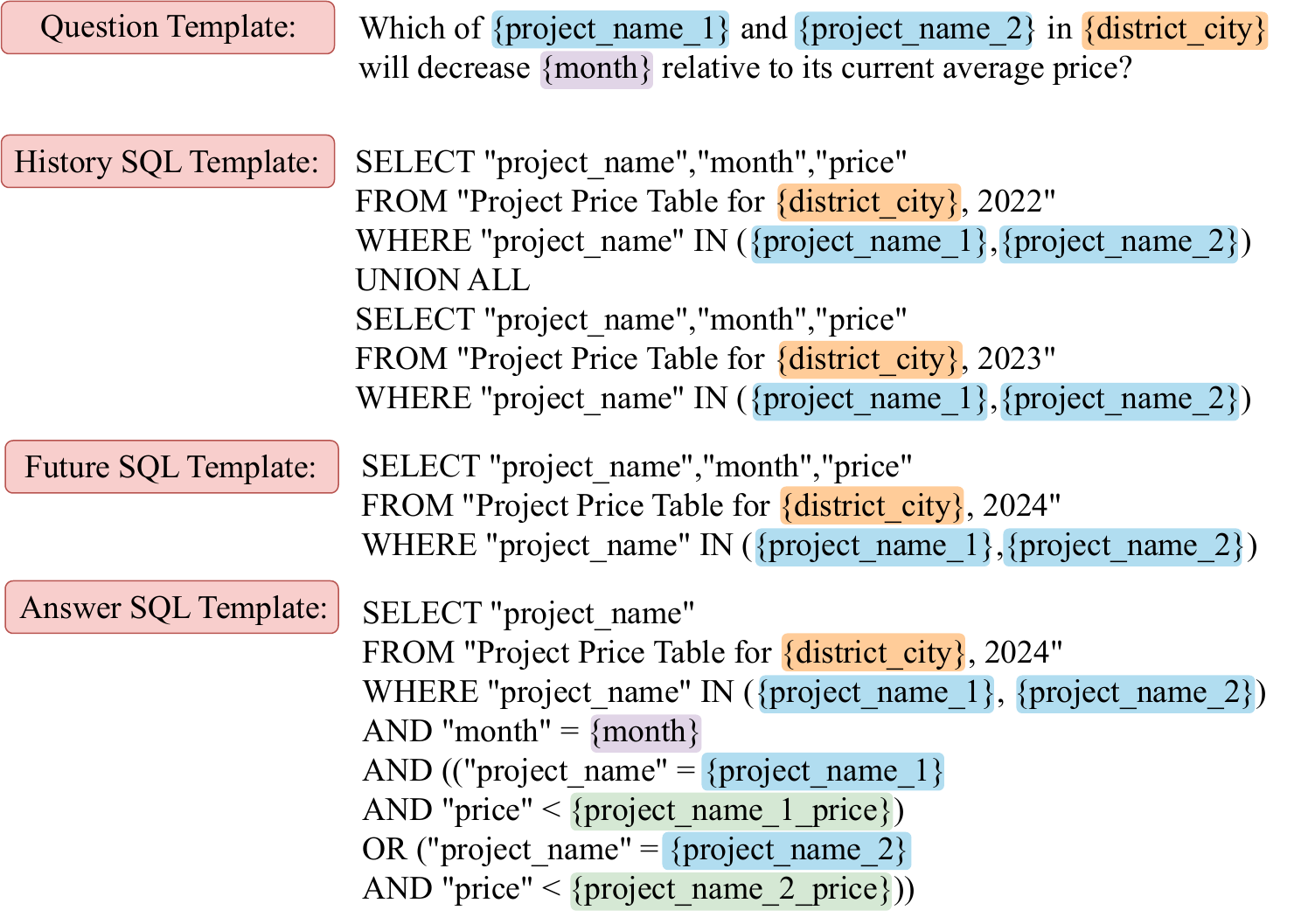}
 \centering
 \caption{Examples of template filling.}
 \label{fig:Template_Filling}
\end{figure}

Template filling is a widely adopted approach for dataset construction in question-answering research. This methodology offers two key advantages: first, it ensures that each generated question has a deterministic and verifiable answer, which is essential for rigorous evaluation; second, it mitigates human errors and subjective biases that may arise during crowd-sourced data collection. Previous works \cite{DBLP:conf/acl/ChenL023, RETQA@aaai2025, DBLP:conf/nips/ChengYFGYK0L24} have successfully employed template filling for dataset generation.

Based on 288 tables, we design 26 seed templates, as shown in Figure \ref{fig:Template_Filling}. These templates comprise 7 time series forecast seed templates (4 for single-project scenarios and 3 for multi-project scenarios) and 19 forecast-based reasoning seed templates (3 for single projects and 16 for multiple projects). Each seed template contains a Question Template, a History SQL Template, a Future SQL Template, and an Answer SQL Template. As illustrated in Figure \ref{fig:Template_Filling}, elements enclosed in ``\{\}'' within the templates indicate variables to be populated, with some variables exhibiting a one-to-one correspondence across the four templates. After randomly sampling appropriate variable values from the database, these values are inserted into the corresponding positions in the templates.

In constructing this dataset, data from 2022 and 2023 serve as historical data, and the tables are organized on a yearly basis. Accordingly, the years in the History SQL Template are fixed to 2022 and 2023, while the year in both the Future SQL Template and the Answer SQL Template is set to 2024. Notably, certain variables highlighted with a green background in the Answer SQL Template (such as ``project\_name\_1\_price'' and ``project\_name\_2\_price'') are determined dynamically during sampling. For example, when sampling a specific template, the ``{district\_city}'' variable is first set. Subsequently, all project names are extracted from the target table, from which eligible project names are randomly selected. If a question involves the ``present,'' the program automatically retrieves the current month’s price for the sampled project name(s) and assigns these actual values to ``project\_name\_1\_price'' and ``project\_name\_2\_price''.

\begin{figure}[t]
 \centering
 \includegraphics[width=\columnwidth]{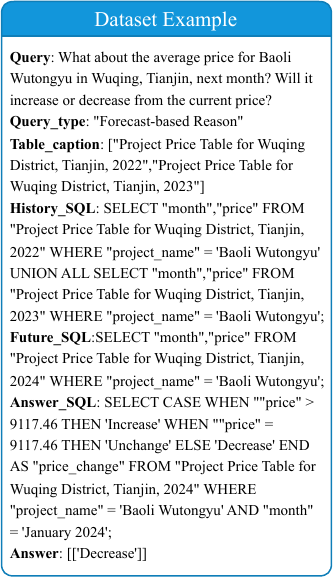}
 \centering
 \caption{A sample of a QA pair.}
 \label{fig:QA_example}
\end{figure}

The question type is identified through fixed phrases embedded within the template. Once all variables have been populated, the SQL statement generated from the filled Answer SQL Template is executed to obtain the answer to the question. An example of a complete Question-Answer (QA) pair is provided in Figure \ref{fig:QA_example}.

\subsection{LLM-based Query Rewriting}
\label{sec:Appendix-LLM-based Query Rewriting}

Template-generated queries often exhibit monolithic syntactic structures, differing significantly from the natural language queries of real users. This discrepancy can lead to suboptimal performance when models trained on such template data process actual user queries. 

To bridge the gap between template-generated rigidity and real-world linguistic diversity, we employ an LLM-based rewriting strategy. This is not merely a stylistic change; it aims to simulate the varied syntax and paraphrasing patterns found in natural user interactions, thereby reducing the structural bias inherent in fixed templates. This rewriting process, predicated on preserving the original query semantics, utilizes strategies such as synonym replacement and syntactic structure adjustments to generate queries with more diverse and natural linguistic expressions. Specifically, we utilized the API of the Qwen2.5-72B model for this query rewriting task, aiming to generate outputs that more closely align with human linguistic patterns. Details of the prompts employed for rewriting are provided in Appendix \ref{sec:Appendix-all_prompts} Figure \ref{fig:Rewrite_prompt}.

To evaluate the rewriting effect, we randomly selected 50 original template queries and their corresponding LLM-rewritten versions. These 100 queries, after being randomly shuffled, were subjectively scored by seven volunteers based on their naturalness (1 to 5 points). The scoring criteria were explicit: the higher the score, the closer the sentence aligns with natural human language; conversely, a lower score indicates more pronounced template-generated characteristics. For each query, we removed the highest and lowest of the seven scores, then averaged the remaining five. We then calculated the mean score separately for the original template queries and the LLM-rewritten queries. The final average score for the original template queries was 3.324, while the LLM-rewritten queries achieved a final average score of 3.936. These results demonstrate that the LLM-rewritten queries more closely resemble natural human language expression.

Furthermore, we manually inspected a random subset of the rewritten queries to ensure semantic consistency with the original template intentions, confirming that the rewriting process improved naturalness without compromising the logical integrity of the questions.

\subsection{Dataset Statistics}
\label{sec:Appendix-Dataset Statistics}

\begin{table}[t]
\begin{tabular}{ll}
\hline
Statistics                             & Num   \\ \hline
\# Total Questions                     & 28507 \\
\# Train                               & 16944 \\
\# Validation                          & 5742  \\
\# Test                                & 5821  \\
\# Ave, Items per Question             & 2.3   \\
\# Ave, Length per Question            & 32.2  \\
\# Time Series Forecasting   Questions & 8042  \\
\# Numerical Reasoning   Questions     & 20465 \\
\# Unique Tables                       & 288   \\
\# Ave, Rows per Table                 & 845.4 \\ \hline
\end{tabular}
\caption{Dataset statics}\label{Dataset statics}
\end{table}

The ODTQA-FoRe dataset comprises 288 tables and 28,507 question-answering pairs, offering a comprehensive benchmark for evaluating complex question-answering systems. Key characteristics include an average of approximately 2.3 query projects per question and an average of 845.4 rows per table. By question type, the dataset consists of 8,042 time-series forecasting questions (28.21\%) and 20,465 numerical reasoning questions (71.79\%). To ensure fair model development and evaluation, the dataset is partitioned into training (16,944 entries), development (5,742 entries), and test sets (5,821 entries). Consequently, its diverse question types, the substantial volume of its tables, and the ambiguous nature of its temporal expressions establish ODTQA-FoRe as a challenging and comprehensive benchmark for Future Data Forecasting and Reasoning over open-domain scenarios. Detailed statistics are presented in Table \ref{Dataset statics}.

\section{Data Flow and Intermediate Representations}
\label{sec:Appendix-Data Flow and Intermediate Representations}

To clarify the exact mechanism of data passing between agents, we detail the serialization and constraints below:

\textbf{Data Serialization from Retriever to Forecaster:}
The Retriever does not pass raw database rows directly. Instead, it formats the SQL execution results into a standardized list of textual triples: \texttt{[project name, year–month, price]}. This format is explicitly defined in the system prompt (see Figure \ref{fig:Forecaster_prompt}).

\textbf{Tool Invocation by Forecaster:}
The Forecaster's LLM receives these textual triples and autonomously preprocesses them into a clean numerical list (history length = 24) to invoke the \texttt{imputationThenPredictionTool}. Missing values are represented as placeholders. The tool returns a 12-month forecast, which the agent formats back into the triple structure.

\textbf{Constraints:}
Each query instance involves a fixed input size: 24 historical months + 12 forecast months + instructions. This results in a total token count significantly lower than the context window limits of the employed LLMs. Therefore, no chunking or truncation strategies are required for the time-series data.

\section{Supplementary experiments}
\label{sec:Appendix-Supplementary experiments}

\subsection{Table retrieval}

\begin{table*}[]\centering
\begin{tabular}{cc|ccc}
\hline
Model & Method & P & R & F1 \\ \hline
\multicolumn{2}{c|}{BM25} & 90.45 & 90.45 & 90.45 \\ \hline
\multirow{2}{*}{Qwen3   30B} & Summary & 94.60 & 94.71 & 94.65 \\
 & Summary+BM25 & 97.72 & 97.72 & 97.72 \\ \hline
\multirow{2}{*}{Qwen3   Next 80B} & Summary & 95.71 & 95.71 & 95.71 \\
 & Summary+BM25 & 97.63 & 97.63 & 97.63 \\ \hline
\multirow{2}{*}{GPT OSS 20B} & Summary & 96.04 & 96.04 & 96.04 \\
 & Summary+BM25 & 98.12 & 98.14 & 98.13 \\ \hline
\multirow{2}{*}{GPT OSS 120B} & Summary & 97.84 & 97.84 & 97.84 \\
 & Summary+BM25 & 99.21 & 99.21 & 99.21 \\ \hline
\multirow{2}{*}{GLM 4.5 AIR} & Summary & 94.19 & 94.19 & 94.19 \\
 & Summary+BM25 & 97.59 & 97.59 & 97.59 \\ \hline
\end{tabular}
\caption{Comparison of table retrieval performance.}\label{Table_caption_search}
\end{table*}

Traditional open-domain tabular question answering typically relies on BM25 to directly match target table captions with the input question. However, our approach first leverages the inductive reasoning and summarization capabilities of large language models (LLMs) to generate a table-caption summary based on the query; if this summary does not correspond to any caption in the database, we then apply BM25 to identify the most relevant table caption. This two-stage strategy not only addresses the uncertainty associated with table count in multi-table scenarios, but also enhances the accuracy of table retrieval. In our dataset, historical queries tend to select the complete sequence of time steps, which results in a fixed number of target tables for each query (namely two, corresponding to two years). 
As shown in Table \ref{Table_caption_search}, our proposed two-stage method, i.e., summarization-then-BM25, significantly outperforms the traditional BM25 direct matching approach, thereby validating the effectiveness of our strategy in this specific scenario.

\subsection{SQL performance}
\label{sec:Appendix-SQL performance}

\begin{table}[th]\centering
\begin{tabular}{c|cc}
\hline
Model & ECR & EA \\ \hline
Qwen3 30B & \textbf{99.85} & 71.38 \\
Qwen3 Next 80B & 93.47 & \textbf{85.72} \\
GPT OSS 20B & 97.62 & 72.34 \\
GPT OSS 120B & 87.84 & 67.93 \\
GLM 4.5 AIR & 94.30 & 71.59 \\ \hline
\end{tabular}
\caption{Comparison of LLM-generated SQL execution results of TimeFore.}\label{SQL_metrics}
\end{table}

To evaluate the performance of SQL generation in the Retriever module, we employ two metrics: Executable Code Ratio (ECR) and Execution Accuracy (EA) \cite{spider@emnlp/YuZYYWLMLYRZR18}. ECR measures the proportion of generated code that can be executed, thus reflecting the model's ability to produce runnable SQL statements. The EA metric represents the percentage of tasks for which the first generated code successfully passes the test case. In our study, since each sample is associated with a single test case designed to query tables, EA is equivalent to accuracy.

As shown in Table \ref{SQL_metrics}, Qwen3 30B achieves the highest ECR for generated SQL, while Qwen3 Next 80B obtains the highest EA score.

\subsection{Time series imputation}
\label{sec:Appendix-Time series imputation}

\begin{table}[t]
\begin{tabular}{c|ccc}
\hline
Model & MSE & MAE & MRE \\ \hline
TimesNet & \textbf{1.78E+07} & \textbf{2169.53} & \textbf{0.0728} \\
i-Transformer & 2.30E+07 & 2720.74 & 0.0926 \\
TimeMixer & 1.85E+07 & 2234.00 & 0.0755 \\ \hline
\end{tabular}
\caption{Performance of three models for time series imputation task.}\label{TimesNet_metrics}
\end{table}

Specialized time-series models typically outperform general large language models in time-series forecasting; however, these models rely heavily on large-scale and standardized datasets. In practical scenarios such as real estate, severe data missingness is common, posing significant challenges for time-series analysis. To address this, we utilize the imputation functionality of the TimesNet model, allowing us to handle missing values and ensure the effectiveness of subsequent prediction tasks.

Notably, both the training of TimesNet for imputation and for forecasting require well-constructed and appropriate datasets. Therefore, building on the project sequence data introduced in Section \ref{data source}, we further construct dedicated datasets for both time-series imputation and time-series forecasting to meet the needs of our study.

For the time-series imputation task, we start by selecting project-level sequences from the training, validation, and test sets that contain at least six months of historical records between 2022 and 2023, specifically targeting historical query data with missing values. The final imputation dataset consists of 8,418 training sequences, 2,815 validation sequences, and 2,853 test sequences.

Subsequently, for the forecasting task, we again use the same project divisions and select sequences with at least nine months of historical data during 2022–2023 and at least two additional months of data in 2024, creating a dataset fit for forecasting future trends. This results in a forecasting dataset comprising 5,806 training sequences, 1,975 validation sequences, and 1,963 test sequences.

Our experimental workflow proceeds as follows: We first train a TimesNet model for time-series imputation according to the official implementation, and apply the trained model to fill in missing values within the forecasting dataset (across all data partitions). We then retrain TimesNet on the now-complete forecasting dataset for the time-series forecasting task. Model performance is evaluated on the test sets of both the imputation and forecasting datasets, as reported in Table \ref{TimesNet_metrics}. When fine-tuning time series forecasting models, we consistently use the optimal hyperparameter settings recommended in the original publications. For scenarios where LLMs perform time series forecasting via few-shot learning, we set the temperature parameter of the model to 0.8 uniformly to ensure consistent and controllable results.

\begin{figure}[t]
 \centering
 \includegraphics[width=\columnwidth]{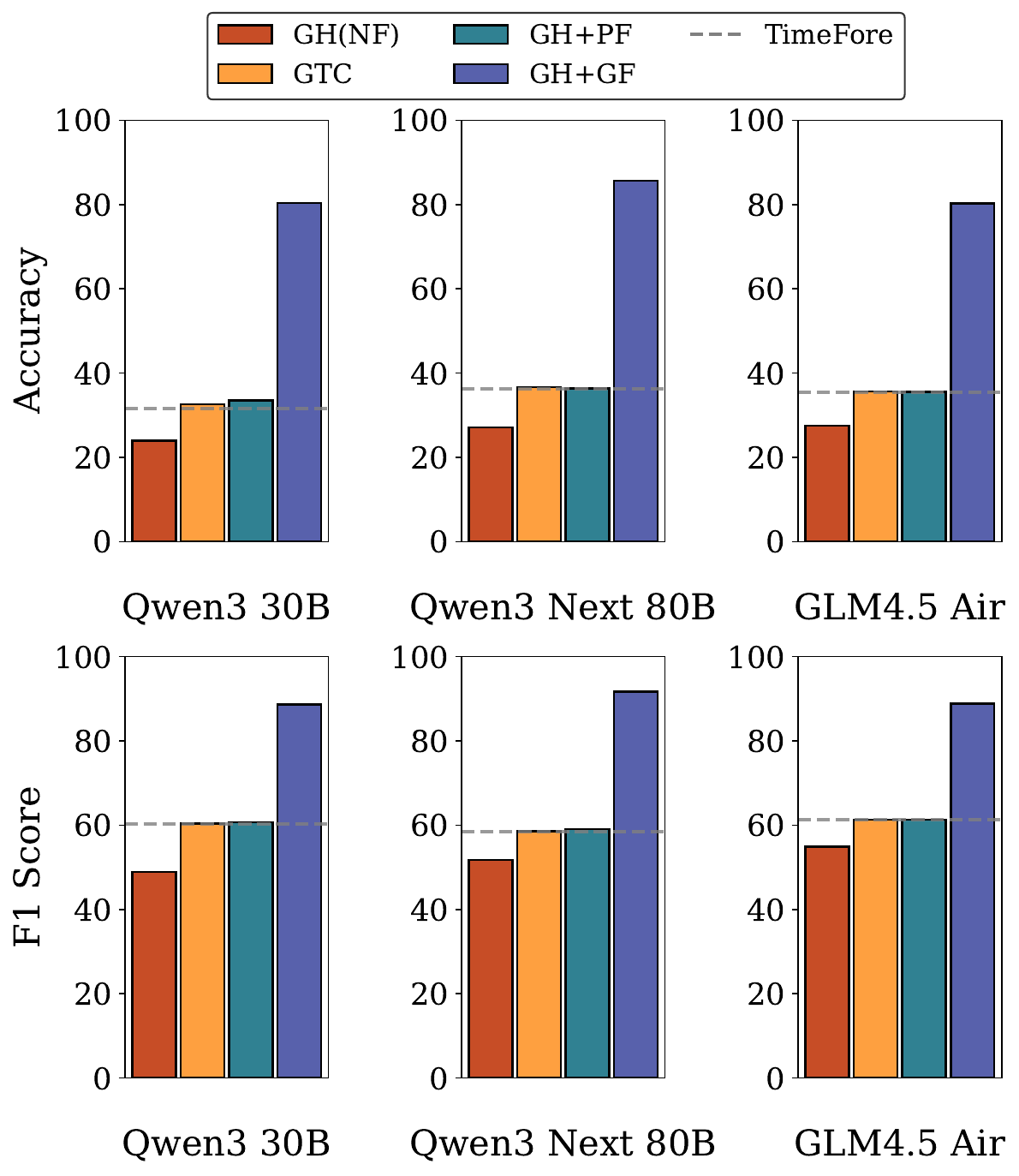}
 \centering
 \caption{Ablation study on the TimeFore framework in forecast-based reasoning tasks.} 
 \label{fig:TSF_Ablation_Study_All}
\end{figure}

\subsection{Error Decomposition and Bottleneck Analysis}
\label{sec:Appendix-Error Decomposition and Bottleneck Analysis}

To precisely identify the source of errors within the TimeFore pipeline, we conduct a quantitative decomposition based on our ablation study results and the performance of individual modules.

Specifically, we attribute the overall error to four main sources: (1) Table Retrieval, (2) Data Acquisition (SQL Generation), (3) Time-series Forecasting, and (4) Final Reasoning. We quantify the impact of each source by measuring the performance drop when the ground truth for that specific stage is replaced with the model's output.

Based on Figure \ref{fig:TSF_Ablation_Study_All}, we observe the following:

\begin{itemize}

\item \textbf{Data Retrieval is Robust}: The limited performance gains observed when providing golden table captions (GTC) or perfect historical data (GH+PF)—with maximum accuracy improvements of only approximately 1.02\% and 1.93\%, respectively—indicate that SQL generation errors are not the primary cause of overall system failures. This finding is further supported by the decoupling between SQL capability and final reasoning performance. For instance, while the Qwen3 Next 80B model achieves the highest SQL EA score (85.72\%) as shown in Table \ref{SQL_metrics}, its Forecast-based Reasoning F1 score (58.41\%) in Table \ref{Overall_performance} is lower than that of the Qwen3 30B model (60.25\%), despite the latter's inferior SQL performance. This discrepancy confirms that precise SQL generation does not necessarily translate to proportional gains in the end-to-end task, suggesting that data retrieval is a robust component and not the core bottleneck.

\item \textbf{Forecasting is the Primary Bottleneck}: Quantitatively, replacing predicted future data with ground truth yields substantial accuracy gains: Qwen3-30B improves by 46.80\%, Qwen3-Next 80B by 49.23\%, and GLM 4.5 Air by 44.73\%. These significant increases confirm that forecasting inaccuracies constitute the primary bottleneck, limiting even the strongest reasoning capabilities.

\item \textbf{Retrieval and Reasoning}: While the Table Retrieval module achieves high F1 scores (above 97.59\%, see Appendix Table \ref{Table_caption_search}), it still contributes a non-negligible portion of errors in open-domain scenarios. Furthermore, the gap between GH+GF and the theoretical upper bound (100\%) reflects the limitations of the LLM in conducting complex numerical reasoning or formatting, which constitutes the smallest but still present error source.

\end{itemize}

In summary, the error analysis confirms that while our Retriever and SQL generation are reliable, the forecasting accuracy of the time-series model remains the critical bottleneck limiting the end-to-end performance of ODTQA-FoRe tasks.

\subsection{Ablation Study of the Analyzer}
\label{sec:Appendix-Ablation Study of the Analyzer}

\begin{table}[]\small
\begin{tabular}{c|c|ccc}
\hline
Model & Method & P & R & F1 \\ \hline
BERT & FT & \textbf{99.96} & \textbf{99.99} & \textbf{99.98} \\
Qwen3 30B & ICL & 99.24 & 99.75 & 99.49 \\
Qwen3 Next 80B & ICL & 99.86 & 99.95 & 99.91 \\
GLM 4.5 Air & ICL & 94.51 & 82.58 & 86.67 \\ \hline
\end{tabular}
\caption{Comparison of the Analyzer module’s binary classification results on input questions, where macro-average is reported and “ICL” and “FT” refer to in-context learning and fine-tuning, respectively.}\label{Query_type_classification_metrics}
\end{table}

In the Analyzer, we categorize queries by type and assign each to a corresponding prompt, thereby enhancing overall system performance. To accomplish this classification, we explore two distinct approaches. The first approach leverages in-context learning (ICL) with a large language model (LLM), enabling direct query type determination using only a handful of examples and eliminating the need for large annotated datasets. The second approach involves fine-tuning a BERT classifier\footnote{https://huggingface.co/google-bert/bert-base-chinese} on query-type annotations drawn from our QA dataset.

As shown in Table \ref{Query_type_classification_metrics}, a fine-tuned BERT model achieves performance comparable to that of the Qwen3 30B and Qwen3 Next 80B models equipped with in-context learning (ICL).
With traditional approaches requiring substantial amounts of labeled data, LLM-based in-context learning offers a highly effective and economical few-shot solution. This makes it particularly well-suited for scenarios where there is insufficient training data or only a few labeled examples are available.

In the evaluation, we tested the numerical extraction module, which aims to standardize answer generation for different query types. As previously discussed, large language models (LLMs) often produce lengthy explanations instead of concise numerical predictions, even for some simple forecasting tasks. To measure answer effectiveness, we introduce the Valid Completion Rate: the proportion of predictions that are both complete and structurally correct, with all target values reliably extracted as required.

\begin{figure}[t]
 \centering
 \includegraphics[width=\columnwidth]{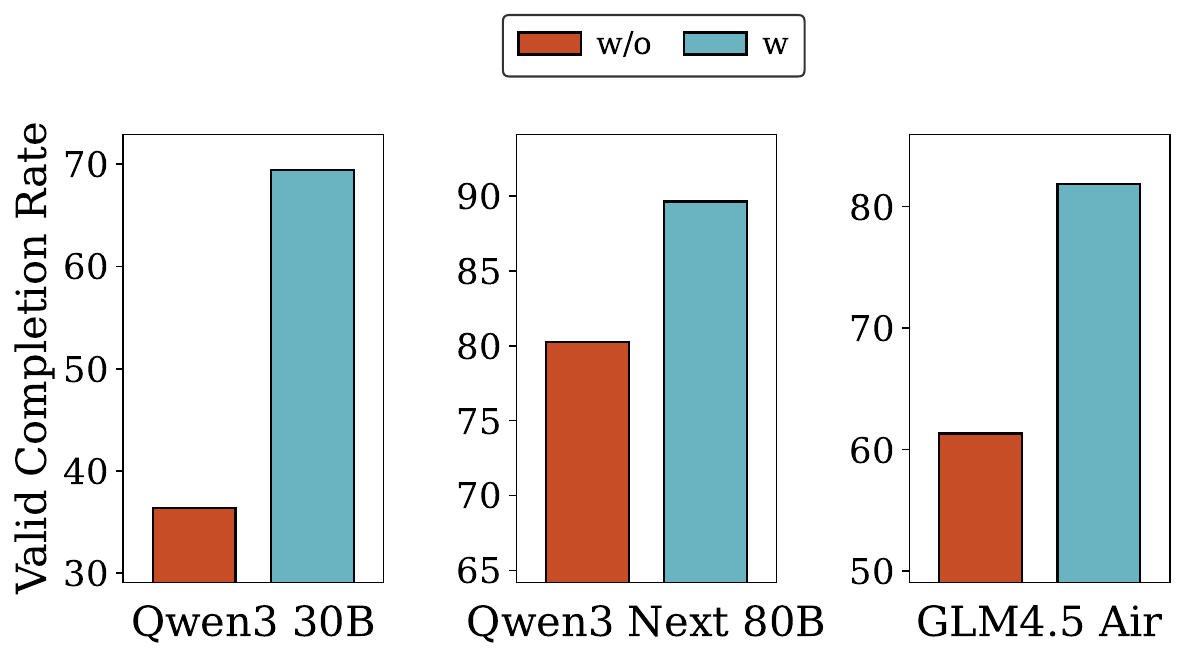}
 \centering
 \caption{Ablation study of the Analyzer on time-series forecasting tasks, where ``w/o'' denotes TimeFore without the numerical extraction module, and ``w'' denotes the complete TimeFore.}
 \label{fig:RSR_Ablation_Study}
\end{figure}

For a fair comparison, other modules remain unchanged—the Retriever continues to retrieve historical data, and the Forecaster provides time series forecasts. The numerical extraction module in the Analyzer is removed, so the model outputs answers directly without post-processing. As shown in Figure \ref{fig:RSR_Ablation_Study}, removing the numerical extraction module decreases the Valid Completion Rate for both models, with Qwen3 30B A3B exhibiting the largest drop (33.08\%). This highlights the crucial role of the numerical extraction module in the Analyzer for improving the effectiveness and reliability of the TimeFore framework.

\section{Computing Infrastructure Statement}
\label{sec:Appendix-Computing Infrastructure Statement}

All neural network models were implemented using PyTorch v2.3.1\footnote{\url{https://pytorch.org/}}. A single NVIDIA GeForce RTX 4090 GPU was utilized for training both the BERT model for query type classification and all time series forecasting models.

For the LLM experiments, we performed inference using the SGLang library\footnote{\url{https://docs.sglang.ai/}} on a cluster of twenty NVIDIA A800-SXM4-80GB GPUs. Specifically, we allocated GPUs as follows: two for Qwen3 30B, four for GLM 4.5-air, four for Qwen3 Next 80B, two for GPT-OSS 20B, and eight for GPT-OSS 120B.

Regarding the implementation of the agent pipeline, we utilized the LangChain framework\footnote{\url{https://www.langchain.com/}} (specifically the LangGraph components) to orchestrate the interactions between the Retriever, Forecaster, and Analyzer agents. To ensure reproducibility and robustness, we adopted the default configurations provided by the framework. Specifically, the temperature parameter for all LLM API invocations and the maximum number of iterations for the agents (e.g., the retry mechanism during SQL generation) were set to the default values provided by LangChain, without manual fine-tuning.

\section{Prompts}
\label{sec:Appendix-all_prompts}

\begin{figure*}[hp]
 \centering
 \includegraphics[width=\textwidth]{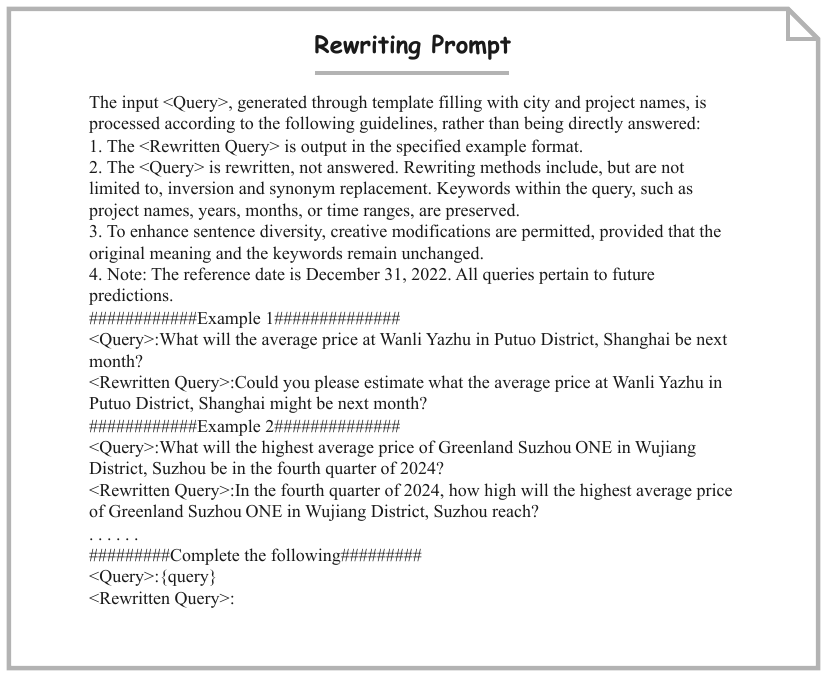}
 \centering
 \caption{Prompt used for rewriting the query generated after template filling.}
 \label{fig:Rewrite_prompt}
\end{figure*}

This section summarizes the prompts employed in data construction and the TimeFore framework, incorporating specific tool-use strategies for robust execution. The rewriting prompt used during data construction is detailed in Figure \ref{fig:Rewrite_prompt}.

\begin{figure*}[hp]
 \centering
 \includegraphics[width=\textwidth]{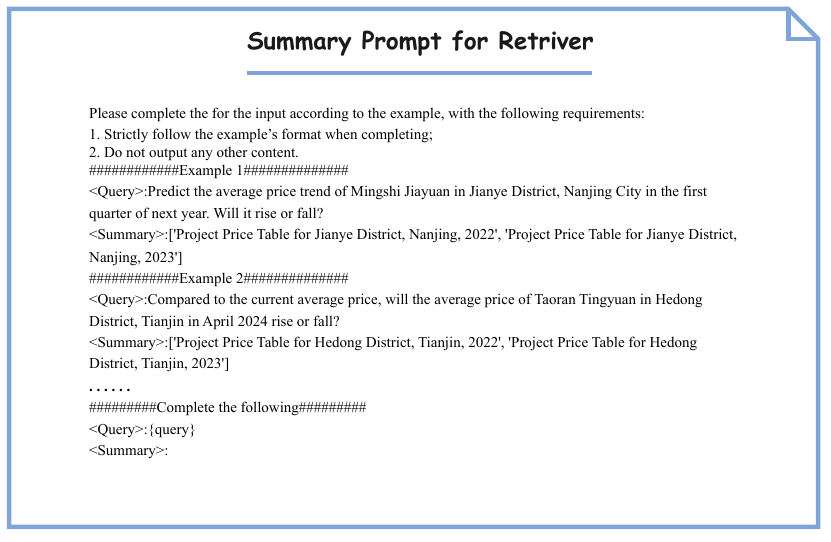}
 \centering
 \caption{Prompt used by LLMs in the Retriever module for query summarization.}
 \label{fig:Summary_prompt}
\end{figure*}

\begin{figure*}[hp]
 \centering
 \includegraphics[width=\textwidth]{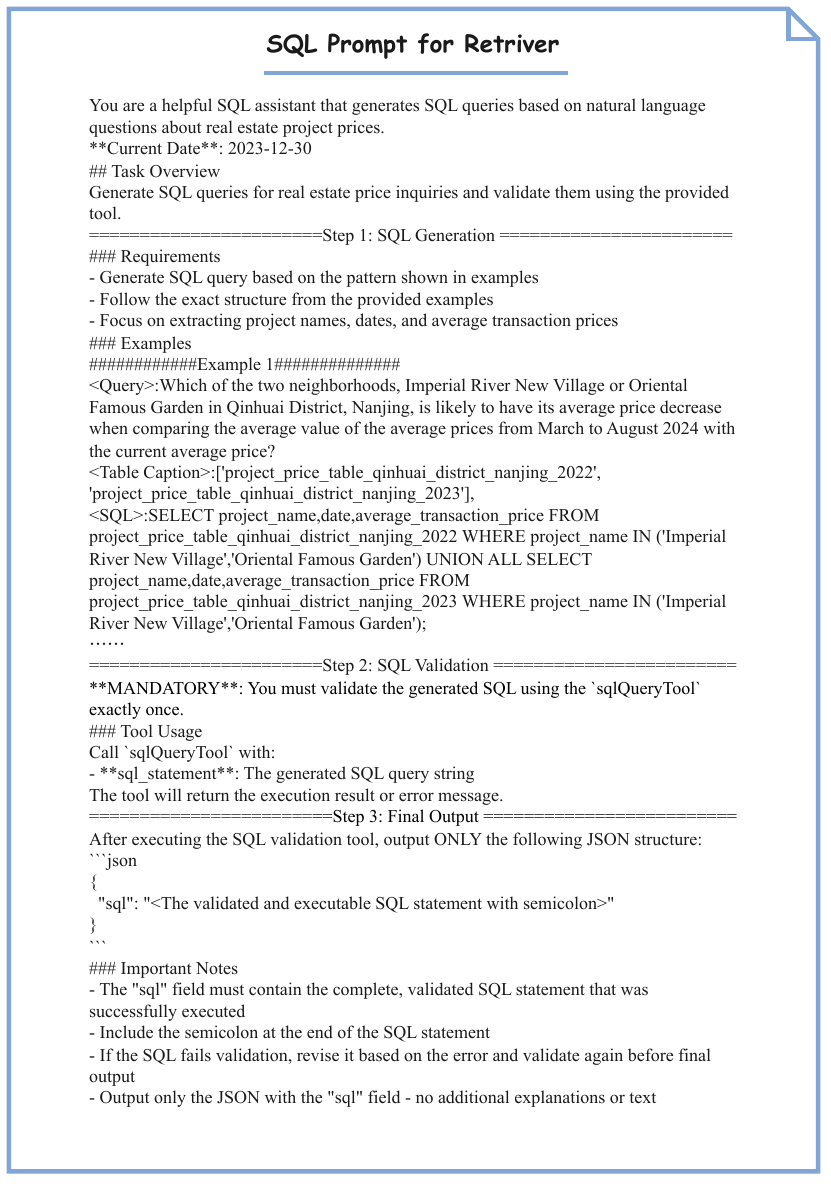}
 \centering
 \caption{Prompt used by LLMs in the Retriever module for SQL generation.}
 \label{fig:SQL_prompt}
\end{figure*}

Within the TimeFore framework, the Retriever module employs a two-stage prompting strategy. First, an LLM summarizes the input query using the prompt in Figure \ref{fig:Summary_prompt}. Subsequently, the SQL generation prompt (Figure \ref{fig:SQL_prompt}) directs the model to formulate a query and immediately verify its validity by executing it via the sqlQueryTool (Algorithm \ref{alg:sqlQuery}). This ensures that only executable SQL statements are passed downstream.

\begin{figure*}[hp]
 \centering
 \includegraphics[width=\textwidth]{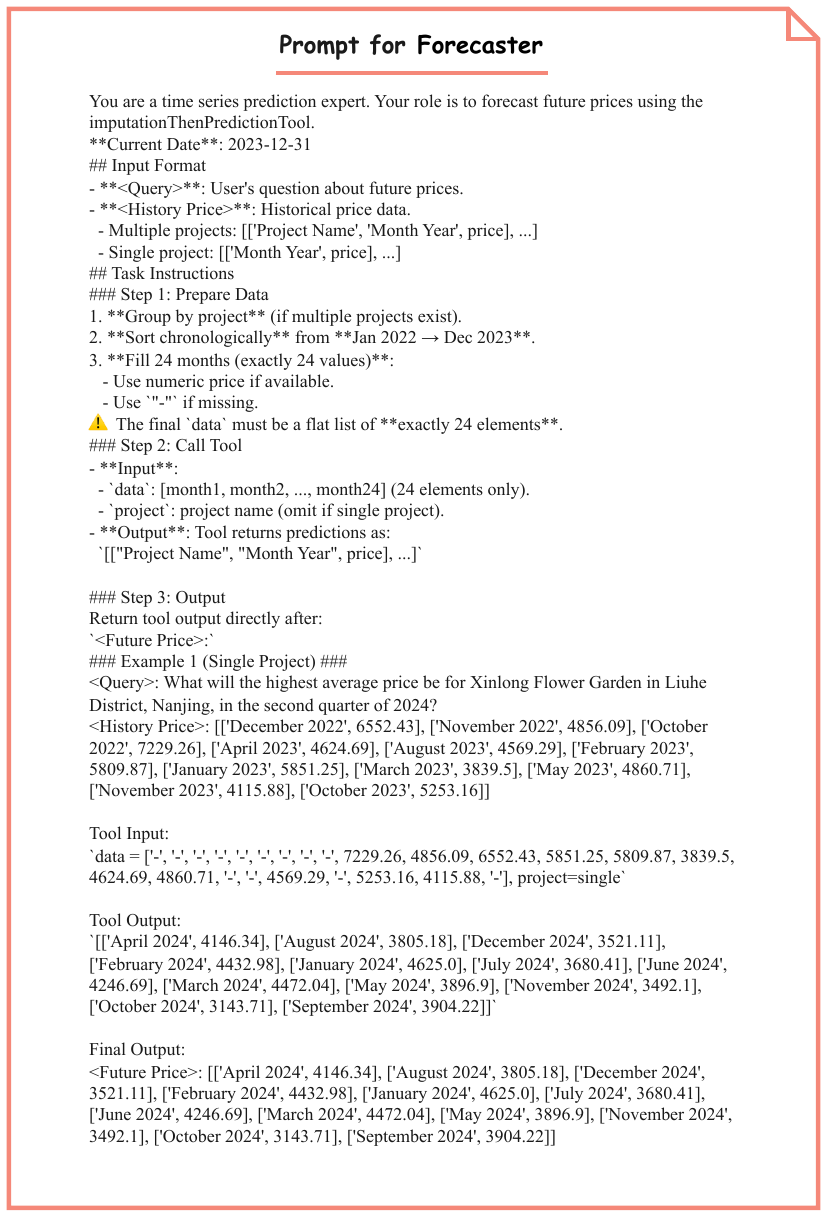}
 \centering
 \caption{Prompt used by LLMs in the Forecaster module for time-series forecasting.}
 \label{fig:Forecaster_prompt}
\end{figure*}

The Forecaster module is designed to handle missing values and prediction simultaneously. It utilizes the prompt shown in Figure \ref{fig:Forecaster_prompt} to invoke the imputationThenPredictionTool (Algorithm \ref{alg:imputationPrediction}), which performs imputation on historical data before generating future time series forecasts.

\begin{figure*}[hp]
 \centering
 \includegraphics[width=\textwidth]{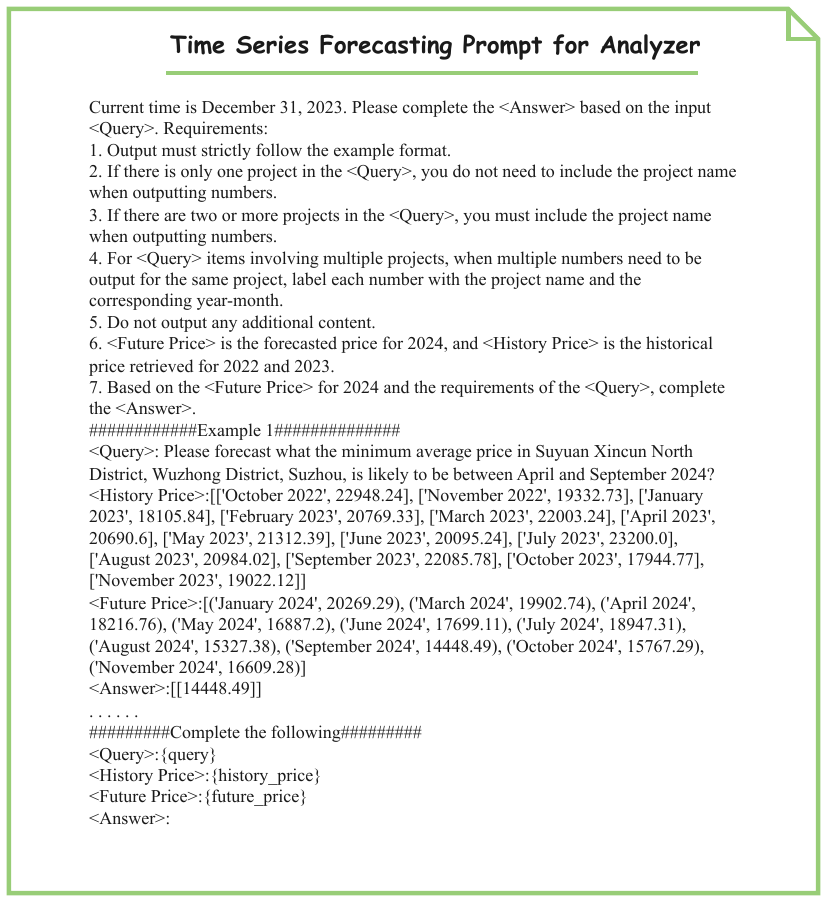}
 \centering
 \caption{Prompt used by LLMs in the Analyzer module for time-series forecasting.}
 \label{fig:RSR_numerical_prompt}
\end{figure*}

\begin{figure*}[hp]
 \centering
 \includegraphics[width=\textwidth]{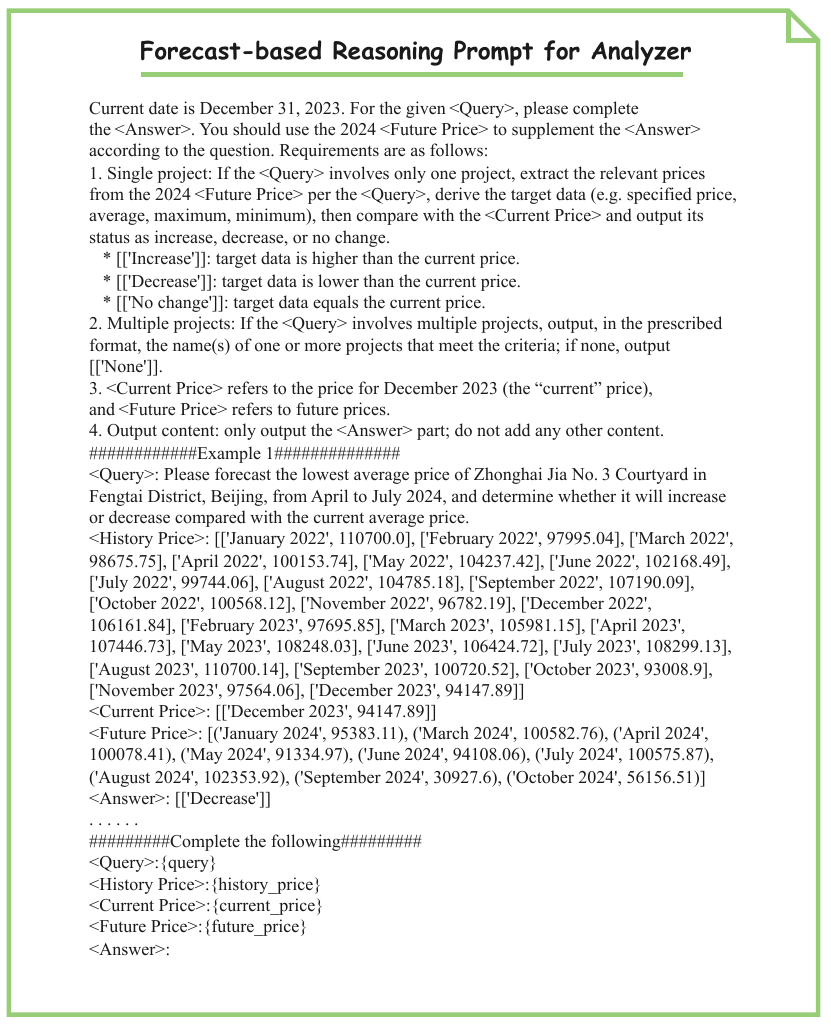}
 \centering
 \caption{Prompt used by LLMs in the Analyzer module for forecast-based reasoning.}
 \label{fig:RSR_judgement_prompt}
\end{figure*}

\begin{figure*}[hp]
 \centering
 \includegraphics[width=\textwidth]{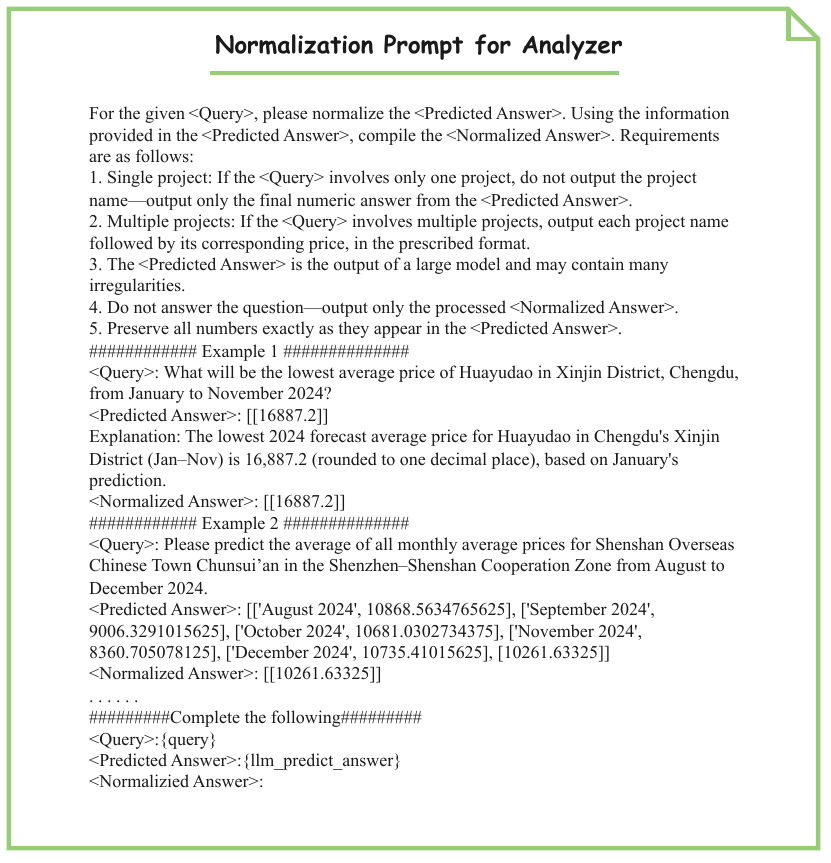}
 \centering
 \caption{Prompt used by LLMs in the Analyzer module for answer normalization.}
 \label{fig:RSR_normalization_prompt}
\end{figure*}

Following forecasting, the Analyzer module selects a distinct prompt based on the query type: the prompt in Figure \ref{fig:RSR_numerical_prompt} is applied for direct time series forecast queries, while forecast-based reasoning queries utilize the prompt in Figure \ref{fig:RSR_judgement_prompt}. Finally, the output normalization process and numerical extraction are guided by the prompt presented in Figure \ref{fig:RSR_normalization_prompt}.

\begin{algorithm*}[hp]
\caption{The sqlQueryTool for Database Interaction}\label{alg:sqlQuery}
\begin{algorithmic}[1]

\REQUIRE $sql\_statement$: A PostgreSQL-compatible SQL query string
\ENSURE $result$: String representation of query results or error message

\STATE \textbf{Step 1: Initialization}
\STATE Initialize $PostgresQueryExecutor$ with database credentials

\STATE \textbf{Step 2: Connect and Execute}
\STATE Attempt to establish database connection and create cursor
\IF{Connection or Execution fails}
    \STATE $result \gets$ "Error executing SQL: " + Exception message
\ELSE
    \STATE Execute $sql\_statement$
    \STATE $rows \gets$ Fetch all results
    \STATE Commit transaction
    \STATE $result \gets$ Convert $rows$ to string
\ENDIF

\STATE \textbf{Step 3: Cleanup}
\STATE Close cursor and database connection (finally)

\STATE \textbf{Step 4: Output}
\RETURN $result$

\end{algorithmic}
\end{algorithm*}

\begin{algorithm*}[hp]
\caption{The imputationThenPredictionTool for Time Series Forecasting}\label{alg:imputationPrediction}
\begin{algorithmic}[1]

\REQUIRE $data$: List of 24 historical values; $project$: str; $device$: str
\ENSURE $month\_price$: List of predicted month-price pairs

\STATE \textbf{Step 1: Validate Input}
\STATE Check $|data| = 24$ and all values are numeric or "-"

\STATE \textbf{Step 2: Data Preprocessing}
\STATE Convert $data$ to tensor, replace "-" with NaN
\STATE Create $target\_mask$ for missing positions

\STATE \textbf{Step 3: Imputation with TimesNet}
\STATE $completed\_inputs \gets$ TimesNet($data$, $target\_mask$)

\STATE \textbf{Step 4: Prediction with TimeXer}
\STATE Prepare decoder inputs and time features
\STATE $predictions \gets$ TimeXer($completed\_inputs$)

\STATE \textbf{Step 5: Format Output}
\STATE Combine month labels with predicted values
\STATE Format as [$project$, $month$, $price$] or [$month$, $price$]

\RETURN $month\_price$

\end{algorithmic}
\end{algorithm*}

\end{document}